\documentclass[final,3p,times]{elsarticle}

\usepackage{lineno,hyperref}
\usepackage{lineno,hyperref}
\usepackage{graphicx}
\usepackage{dcolumn}
\usepackage{bm}
\usepackage{float}
\usepackage{epsfig}
\usepackage{booktabs}
\usepackage{subfigure}
\usepackage{graphics}
\usepackage{amssymb}
\usepackage{amsmath}
\usepackage{array}
\usepackage{color}
\usepackage{booktabs}
\usepackage{multirow}
\usepackage{setspace}
\usepackage{nomencl}
\usepackage{bookmark}
\usepackage{cases}
\usepackage{tabularx}
\usepackage{extarrows}
\usepackage{xcolor}
\usepackage{algorithm}
\usepackage{algpseudocode}
\usepackage[utf8]{inputenc}

\makenomenclature

\usepackage{graphicx}

\modulolinenumbers[5]

\begin{document}

\begin{frontmatter}

\title{A unified MRT-LB framework for Navier-Stokes and nonlinear convection-diffusion equations and beyond: moment equations, auxiliary moments, multispeed lattices, and Hermite matrices}

\author[mymainaddress]{Baochang Shi\corref{mycorrespondingauthor}}
\cortext[mycorrespondingauthor]{Corresponding author}
\ead{shibc@hust.edu.cn}
\author[mysecondaryaddress]{Xiaolei Yuan}
\author[mymainaddress,mythirdaryaddress,myfourtharyaddress]{Zhenhua Chai}

\address[mymainaddress]{School of Mathematics and Statistics, Huazhong University of Science and Technology, Wuhan 430074, China}
\address[mysecondaryaddress]{College of Mathematics and Information Science, Hebei University, Baoding 071002, China}
\address[mythirdaryaddress]{Institute of Interdisciplinary Research for Mathematics and Applied Science, Huazhong University of Science and Technology, Wuhan 430074, China}
\address[myfourtharyaddress]{Hubei Key Laboratory of Engineering Modeling and Scientific Computing, Huazhong University of Science and Technology, Wuhan 430074, China}

\begin{abstract}
We develop a unified multi-relaxation-time lattice Boltzmann (MRT-LB) framework based on discrete Hermite polynomials (Hermite matrices) for the Navier-Stokes equations (NSEs) and nonlinear convection-diffusion equations (NCDEs), using multispeed rectangular lattice (rD$d$Q$b$) models. For NSEs, the proposed MRT-LB model simulates incompressible and compressible isothermal flows in both single-phase and multiphase systems. Macroscopic moment equations are derived from the MRT-LB model via the direct Taylor expansion method. By selecting appropriate fundamental moments, the target NSEs and NCDE are recovered from these moment equations. Critically, the elimination of spurious terms and/or the recovery of the desired terms relies on specific auxiliary moments: the second-order auxiliary moment ($\mathbf{M}_{2G}$) of the source term distribution function (SDF) and the third-order auxiliary moment ($\mathbf{M}_{30}$) of the equilibrium distribution function (EDF) for NSEs, as well as the first-order auxiliary moment ($\mathbf{M}_{1G}$) of the SDF and the second-order auxiliary moment ($\mathbf{M}_{20}$) of the EDF for NCDE. Furthermore, using the weighted orthogonality of Hermite matrices, we establish essential relations for weight coefficients and construct several multispeed rectangular lattice models, including rD2Q25 and rD3Q53, with subgroup models rD2Q21, rD2Q17, rD2Q13, rD3Q45, and rD3Q33. A generalized third-order equilibrium distribution function is derived. We emphasize that for rectangular lattices, specific elements of the Hermite matrix corresponding to third-order discrete Hermite polynomials require correction to satisfy weighted orthogonality.
\end{abstract}


\end{frontmatter}

\linenumbers

\section{Introduction}

Originating from the lattice gas automata in the 1980s, the lattice Boltzmann method (LBM) has undergone over three decades of development. LBM is distinguished by its clear physical foundation, easy in handling boundaries, straightforward implementation, and excellent parallel scalability. Consequently, it has advanced rapidly in fluid science and engineering applications, becoming an important tool for simulating complex fluid systems with wide applications in heat and mass transfer, multiphase flow, porous media flow, and chemical reaction flow, etc. \cite{Chen1998,Succi2001,Guo2013,Kruger2017}.

Many distinct LB models have been developed over the past three decades, including the commonly used single-relaxation-time LB (SRT-LB) model (or lattice Bhatnagar-Gross-Krook (LBGK) model) \cite{Qian1992, Chen1998}, the two-relaxation-time LB (TRT-LB) model \cite{Ginzburg2005, Ginzburg2008}, and the multiple-relaxation-time lattice Boltzmann (MRT-LB) model \cite{dHumieres1992, Lallemand2000, dHumieres2002, Pan2006, Geier2006, Chai2016a}. Most are special forms of  the MRT-LB model \cite{dHumieres2002,ChaiShi2020}. As a more general LB formulation, the MRT-LB collision process uses a matrix to enable decoupled relaxation of different moments, extending the SRT-LB model while improving the numerical stability and accuracy. Pervious studies have shown MRT-LB model's superiority over SRT-LB model in accuracy and stability, with only minor compromises in computational efficiency \cite{dHumieres2002,Luo2011}.

Although diverse LB models exist, most rely on low-Mach-number assumptions and fail to fully recover the energy equations, limiting LBM to isothermal, low-velocity, weakly compressible flows or incompressible flows. To address these limitations, researchers have pursued two strategies for compressible and thermodynamic flows. The first introduces an additional energy distribution function to satisfy the energy equation constraints \cite{Shan1997,Shi2004,Li2007}. For example, Prasianakis
and Karlin \cite{Prasianakis2007,Prasianakis2008} proposed an LB model with correction terms in the kinetic equations for  compressible flows on standard lattices. Li et al. \cite{Li2012} extended Guo et al.'s \cite{Guo2007} double-distribution-function model to weakly compressible flows at the low Mach numbers using diffusion scaling for computing correction terms. However, this strategy increases model complexity due to the added energy distribution function. The second strategy constructs a higher order equilibrium distribution function (EDF), incorporating additional coefficients and multispeed (or multi-layer velocity) lattices to meet additional energy equation constraints. Please refer to the works of Alexander et al. \cite{Alexander1993}, Kataoka et al. \cite{Kataoka2004}, Watari et al. \cite{Watari2007} for the details. In these models, researchers typically preset  discrete equilibrium forms and velocity sets, and then determine coefficients empirically-a process lacking universality. Shan et al. \cite{Shan2006,Shan2016} developed a Hermite-polynomial-expansion based LB model, making equilibrium expansions and the discrete velocity models deterministic rather than empirical.

Notably, most existing models use square lattices and SRT collision operators. Thus, we intent to develop a multispeed rectangular MRT-LB (RMRT-LB) model for NCDE and compressible/incompressible isothermal NSEs. Furthermore, conventional MRT models rely on the specified lattice structure or discrete velocity set and the collision process is carried out in the moment space rather than the velocity (or distribution function) space, which makes the analysis method (e.g., the Chapman-Enskog expansion) very complicated. By contrast, Chai et al. \cite{ChaiShi2020} established a unified MRT-LB framework through velocity-space modeling, introducing a block-lower-triangular relaxation matrix and an auxiliary source distribution function. Recently, they extended this to a general RMRT-LB version  \cite{ChaiShi2023}, deriving a universal rectangular equilibrium distribution function.

\begin{figure}[ht]
\centering
\includegraphics[width=4.5in]{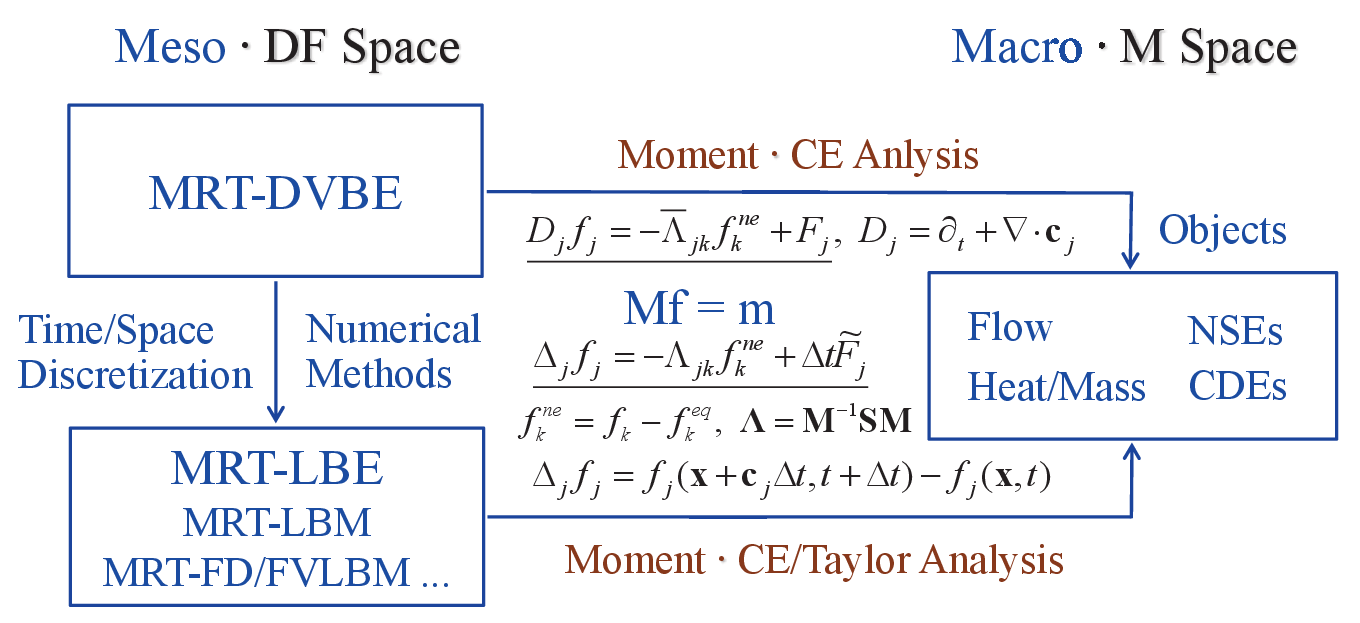}
\caption{\label{fig1}Modeling and elements of MRT-DVBE/LBE, where $f$, $f^{eq}$, $\mathbf M$, $m$, $\Omega$, $\mathbf F$ and $\Lambda$ are the  distribution function (DF), equilibrium, transform matrix, moments, collision term, source DF and collision matrix, respectively.
}
\end{figure}

The MRT-LBM can be derived from the MRT-discrete-velocity Boltzmann equation (MRT-DVBE). Temporal and spatial discretization of MRT-DVBE yields the MRT lattice Boltzmann equation (MRT-LBE), including classical LBM, finite-difference LBM (FD-LBM), and finite-volume LBM (FV-LBM, including its variant, the discrete unified gas kinetic scheme \cite{Guo2013b}). In this work, we won't consider the FD-LBM and FV-LBM. Two paths exist to derive the macroscopic equations (NSEs/NCDEs) from mesoscopic MRT-DVBE or MRT-LBE (see Fig. {\ref{fig1}}): direct moment computation with Chapmann-Enskog (C-E) analysis for MRT-DVBE, or direct Taylor expansion/C-E analysis for MRT-LBE. Building on this, we establish a unified MRT-LBM framework based on the MRT-DVBE or MRT-LBE via mesoscopic direct discrete modelling (DDM) in the velocity space. The unified framework is a direct generalisation of Chai et al.'s model \cite{ChaiShi2023}. Several multispeed lattice models on rectangular lattice, including rD2Q25, rD2Q17,  rD2Q13, rD3Q53 and rD3Q33 lattice models, are constructed based on weighted orthogonality of the zeroth- to third-order Hermite matrices \{$\mathbf{H}_0$, $\mathbf{H}_1$, $\mathbf{H}_2$, $\mathbf{H}_3$\}.

The rest of this paper is organized as follows. Section II presents the RMRT-LB method. Section III derives moment equations of the proposed RMRT-LB method via direct Taylor expansion analysis. Section IV describes the general representation of equilibrium, auxiliary and source distribution functions based on the Hermite matrix. Section V details multispeed lattice models of the RMRT-LB method on rectangular lattice. Some conclusions are summarized in Section VI.

\section{Rectangular Multiple-relaxation-time lattice Boltzmann method}

The evolution equation of the RMRT-LB method with the multispeed rectangular D$d$Q$b$ (rD$d$Q$b$) lattice has the same form as that of the MRT-LB method  in Ref. \cite{ChaiShi2020} and the RMRT-LB method in Ref. \cite{ChaiShi2023}:
\begin{equation}\label{eq:2-1}
 f_j(\mathbf{x}+\mathbf{c}_j \Delta t,t+\Delta t)=f_j(\mathbf{x}, t)-\mathbf{\Lambda}_{jk} f_k^{ne}(\mathbf{x}, t)+\Delta t \big[G_j(\mathbf{x},t)+F_j(\mathbf{x}, t)+\frac{\Delta t}{2}\bar{D}_j F_j(\mathbf{x}, t)\big],
\end{equation}
where $f_j(\mathbf{x}, t)$ is the distribution function at position $\mathbf{x}$ in $d$-dimensional space and time $t$ along the velocity $\mathbf{c}_j$, $f_j^{ne}(\mathbf{x}, t)=f_j(\mathbf{x}, t)-f_j^{eq}(\mathbf{x}, t)$ is the nonequilibrium distribution function (NEDF), and $f_j^{eq}(\mathbf{x}, t)$ is the equilibrium distribution function. $F_j(\mathbf{x}, t)$ is the distribution function of a source or forcing term, $G_j(\mathbf{x}, t)$ is the \textit{auxiliary} distribution function, and $\mathbf{\Lambda}=(\mathbf{\Lambda}_{jk})$ is a $b \times b$ invertible collision matrix. $\Delta t$ is the time step, $\bar{D}_j=\theta\partial_t +\gamma \mathbf{c}_j\cdot \nabla$ with $\theta\geq 0 $, and commonly $\theta =1$ being set, while $\gamma =1$ for NSEs, and $\gamma \geq 0$ for NCDE and usually $\gamma\in \{0,1\}$ . In the evolution equation (\ref{eq:2-1}), the key elements, $\mathbf{c}_j, f_{j}^{eq}, F_j, G_j$ and $\mathbf{\Lambda}$, must be given properly.

The unknown macroscopic conserved variable(s), $\phi(\mathbf{x},t)$ for NCDE, or $\rho(\mathbf{x},t)$ and $\mathbf{u}(\mathbf{x},t)$ for NSEs, can be computed by

\begin{subequations}\label{eq:2-4-0}
\begin{equation}
\phi(\mathbf{x}, t)=\sum_j f_j(\mathbf{x}, t)+\frac{\Delta t \lambda}{2} S_G(\mathbf{x}, t),
\end{equation}
\begin{equation}
\rho(\mathbf{x}, t)=\sum_j f_j(\mathbf{x}, t) +\frac{\Delta t \lambda_1}{2} S_G(\mathbf{x}, t),
\mathbf{u}(\mathbf{x}, t)=\left[\sum_j \mathbf{c}_j f_j(\mathbf{x}, t)+\frac{\Delta t \lambda_2}{2} \mathbf{F}_G(\mathbf{x}, t)\right]/\rho(\mathbf{x}, t),
\end{equation}
\end{subequations}
where $\lambda,\lambda_1$ and $\lambda_2$ are parameters which are used for correcting the source or \emph{partial} source terms, $S_G$ and $\mathbf{F}_G$ for NCDE and NSEs. They will be given later. It can be found that the computation of the unknown macroscopic conserved variable(s) in Eq. (\ref{eq:2-4-0}) is different from that in Refs. \cite{ChaiShi2020} and \cite{ChaiShi2023}. Here, the correction is used. This makes the RMRT-LB model more universal. In fact, the RMRT-LB model (\ref{eq:2-1}) with Eq. (\ref{eq:2-4-0}) contains two types of common LB models as its special cases, namely, the model without correction (set $S_G=0 $ and $ \mathbf{F}_G=\mathbf{0}$ in Eq. (\ref{eq:2-4-0})) \cite{ChaiShi2020, ChaiShi2023}, and the model with correction (set $\theta=0 $ and $ \gamma=0 $ in Eq. (\ref{eq:2-1})) \cite{GuoZhengShi2002}. Furthermore, Eq. (\ref{eq:2-4-0}) contains also the case that only \emph{partial} correction of the source term is required for some problems, as shown in Ref. \cite{RongShi2015}.

The evolution equation (\ref{eq:2-1}) can be divided into two sub-steps, i.e., collision and propagation,
\begin{subequations}\label{eq:2-2}
\begin{equation}
\textbf{Collison:}\ \  \tilde{f}_j(\mathbf{x}, t)=f_j(\mathbf{x}, t)-\mathbf{\Lambda}_{jk} f_k^{ne}(\mathbf{x}, t)+\Delta t \big[G_j(\mathbf{x},t)+F_j(\mathbf{x}, t)+\frac{\Delta t}{2}\bar{D}_j F_j(\mathbf{x}, t)\big],  \\
\end{equation}
\begin{equation}
\textbf{Propagation:}\ \  f_j(\mathbf{x}+\mathbf{c}_j \Delta t,t+\Delta t)=\tilde{f}_j(\mathbf{x}, t), \ \ \ \ \ \ \ \ \ \ \ \ \ \ \ \ \ \ \ \ \ \ \ \ \ \ \ \ \ \ \ \ \ \ \ \ \ \ \ \ \ \ \ \ \ \ \ \ \ \ \ \ \ \ \ \ \ \ \ \
\end{equation}
\end{subequations}
where $\tilde{f}_j(\mathbf{x}, t)$ is the post-collision distribution function.


In the implementation of the RMRT-LB method, one can use two schemes to discretize the term $\bar{D}_j F_j(\mathbf{x}, t)$ on the right hand side of Eq. (\ref{eq:2-1}). Actually, if $\gamma=0$ and $\theta >0$, the first-order explicit difference scheme $\partial_{t} F_j(\mathbf{x}, t)=[F_j(\mathbf{x}, t)-F_j(\mathbf{x}, t-\Delta t)]/\Delta t$ is adopted for NCDEs  \cite{Shi2009,Chai2016a}. For this case the MRT-LB model is a \emph{three level scheme}. For the case of $\gamma=1$ and $\theta=1$, however, we can use the first-order implicit difference scheme $(\partial_{t}+\mathbf{c}_{j}\cdot\nabla) F_j(\mathbf{x}, t)=[F_j(\mathbf{x}+\mathbf{c}_{j}\Delta t, t+\Delta t)-F_j(\mathbf{x}, t)]/\Delta t$ for both NCDE and NSEs, and take the  transform $\bar{f}_j=f_j-\frac{\Delta t}{2}F_j$ as in Refs. \cite{He1998, Du2006,Chai2016a}, then Eq. (\ref{eq:2-1}) becomes the following model with \emph{full} correction \cite{GuoZhengShi2002},
\begin{equation}\label{eq:2-3}
 \bar{f}_j(\mathbf{x}+\mathbf{c}_j \Delta t,t+\Delta t)=\bar{f}_j(\mathbf{x}, t)-\mathbf{\Lambda}_{jk} \bar{f}_k^{ne}(\mathbf{x}, t)+\Delta t \big[G_j(\mathbf{x},t)+(\delta_{jk}-\mathbf{\Lambda}_{jk}/2)F_k(\mathbf{x}, t)\big],
\end{equation}
where $\bar{f}_j^{ne}(\mathbf{x}, t)=\bar{f}_j(\mathbf{x}, t)-f_j^{eq}(\mathbf{x}, t)$. Additionally, we also have the following relations \cite{Ladd1994,Ginzburg1994,Kuzmin2011},
\begin{subequations}\label{eq:2-4}
\begin{equation}
\sum_j f_j(\mathbf{x}, t)= \sum_j\bar{f}_j(\mathbf{x}, t)+\frac{\Delta t}{2}\sum_j F_j(\mathbf{x}, t)= \sum_j\bar{f}_j(\mathbf{x}, t)+\frac{\Delta t}{2} S_F(\mathbf{x}, t),
\end{equation}
\begin{equation}
\sum_j \mathbf{c}_j f_j(\mathbf{x}, t)= \sum_j \mathbf{c}_j\bar{f}_j(\mathbf{x}, t)+\frac{\Delta t}{2}\sum_j \mathbf{c}_j F_j(\mathbf{x}, t)= \sum_j \mathbf{c}_j\bar{f}_j(\mathbf{x}, t)+\frac{\Delta t}{2} \mathbf{F}_F(\mathbf{x}, t).
\end{equation}
\end{subequations}

It follows from Eqs. (\ref{eq:2-4-0}) and (\ref{eq:2-4}) that for NCDE,

\begin{equation}\label{eq:2-4-1}
\phi(\mathbf{x}, t)= \sum_j\bar{f}_j(\mathbf{x}, t)+\frac{\Delta t}{2}(\lambda S_G + S_F)(\mathbf{x}, t),
\end{equation}
or for NSEs,
\begin{subequations}\label{eq:2-4-2}
\begin{equation}
\rho(\mathbf{x}, t)= \sum_j\bar{f}_j(\mathbf{x}, t)+\frac{\Delta t}{2}(\lambda_1 S_G + S_F)(\mathbf{x}, t),
\end{equation}
\begin{equation}
\mathbf{u}(\mathbf{x}, t)= \big[\sum_j \mathbf{c}_j\bar{f}_j(\mathbf{x}, t)+\frac{\Delta t}{2}(\lambda_2 \mathbf{F}_G + \mathbf{F}_F)(\mathbf{x}, t)\big]/\rho(\mathbf{x}, t).
\end{equation}
\end{subequations}

It can be found that the MRT-LB model (\ref{eq:2-3}) with Eq. (\ref{eq:2-4-1}) or Eq. (\ref{eq:2-4-2}) is a \emph{two level implicit scheme} if the correction term is implicit.

\section{The moment equations of RMRT-LB method: Direct Taylor expansion}
\label{printlayout}

Although there are four basic analysis methods that can be used to recover the macroscopic NSEs and NCDE from the LB models, i.e., the Chapman-Enskog (CE) analysis \cite{Chapman1970,Frisch1987,Ginzburg2013}, the Maxwell iteration (MI) method \cite{Ikenberry1956,Yong2016}, the direct Taylor expansion (DTE) method \cite{Holdych2004,Wagner2006,Kaehler2013} and the recurrence equations (RE) method \cite{Ginzburg2012,dHumieres2009,Ginzburg2015}, they all yield the same equations at the second-order of expansion parameters, and the DTE method is much simpler, as shown in Ref. \cite{ChaiShi2020}. In what follows, the DTE method is used to analyze the RMRT-LB model.

Applying the Taylor expansion to Eq. (\ref{eq:2-1}),  one can get
\begin{equation}\label{eq:4-1}
\sum_{l=1}^{N}\frac{\Delta t^l}{l!}D_j^l f_j +O(\Delta t^{N+1})=-\mathbf{\Lambda}_{jk} f_k^{ne}+\Delta t \tilde{F}_j, N\geq 1,
\end{equation}
where $\tilde{F}_j=G_j+F_j+\Delta t \bar{D}_j F_j/2$.

Based on $f_j=f_j^{eq}+f_j^{ne}$ and Eq. (\ref{eq:4-1}), the following equations can be obtained,
\begin{subequations}\label{eq:4-2}
\begin{equation}\label{eq:4-2a}
f_j^{ne}=O(\Delta t),
\end{equation}
\begin{equation}\label{eq:4-2b}
\sum_{l=1}^{N-1}\frac{\Delta t^l}{l!}D_j^l (f_j^{eq}+f_j^{ne}) + \frac{\Delta t^N}{N!}D_j^N f_j^{eq}=-\mathbf{\Lambda}_{jk} f_k^{ne}+\Delta t \tilde{F}_j+O(\Delta t^{N+1}), N\geq 1.
\end{equation}
\end{subequations}
Then from Eq. (\ref{eq:4-2b}), we can derive the equations at first and second orders of $\Delta t$,

\begin{subequations}\label{eq:4-3}
\begin{equation}\label{eq:4-3a}
D_j f_j^{eq}=-\frac{\mathbf{\Lambda}_{jk}}{\Delta t } f_k^{ne}+G_j+F_j+O(\Delta t),
\end{equation}
\begin{equation}\label{eq:4-3b}
D_j (f_j^{eq}+f_j^{ne})+\frac{\Delta t}{2}D_j^2 f_j^{eq}=-\frac{\mathbf{\Lambda}_{jk}}{\Delta t } f_k^{ne}+ G_j+F_j+\frac{\Delta t}{2} \bar{D}_jF_j+O(\Delta t^2).
\end{equation}
\end{subequations}
According to Eq. (\ref{eq:4-3a}), we have
\begin{equation}\label{eq:4-4}
\frac{\Delta t}{2}D_j^2 f_j^{eq}=-\frac{1}{2} D_j \mathbf{\Lambda}_{jk} f_k^{ne}+\frac{\Delta t}{2}D_j(G_j+F_j)+O(\Delta t^2),
\end{equation}
Substituting Eq. (\ref{eq:4-4}) into Eq. (\ref{eq:4-3b}), one can obtain the following equation,
\begin{equation}\label{eq:4-5}
D_j f_j^{eq}+D_j\big(\delta_{jk}-\frac{\mathbf{\Lambda}_{jk}}{2}\big)f_k^{ne}+\frac{\Delta t}{2}D_j (G_j+F_j) =-\frac{\mathbf{\Lambda}_{jk}}{\Delta t} f_k^{ne}+G_j+F_j+\frac{\Delta t}{2} \bar{D}_j F_j+O(\Delta t^2).
\end{equation}

Based on Eqs. (\ref{eq:4-3a}) and (\ref{eq:4-5}), the related macroscopic equation (NSEs and NCDE) can be recovered with some proper constraints on the collision matrix $\mathbf{\Lambda}$ and the moments of $f_j^{eq}$, $G_j$ and $F_j$. For NSEs, if we take $\theta=\gamma=1$, then Eq. (\ref{eq:4-5}) can be simplified by
\begin{equation}\label{eq:4-6}
D_j f_j^{eq}+D_j\big(\delta_{jk}-\frac{\mathbf{\Lambda}_{jk}}{2}\big)f_k^{ne}+\frac{\Delta t}{2}D_j G_j =-\frac{\mathbf{\Lambda}_{jk}}{\Delta t} f_k^{ne}+G_j+F_j+O(\Delta t^2).
\end{equation}

\subsection{The derivation of the moment equations for Navier-Stokes equations using DTE method}

In this subsection, based on Eq. (\ref{eq:4-3}a) and Eq. (\ref{eq:4-5}), we first derive the general moment equations with different time step orders for NSEs from MRT-LB method (\ref{eq:2-1}). Then, by selecting specific moments, the target NSEs can be recovered from these moment equations. The similar discussion on NCDE can be found in Appendix A.

The basic moments of $\mathbf{\Lambda}$, $f_j$, $f_j^{eq}$, $G_j$, and $F_j$ are given as follows,
\begin{subequations}\label{eq:M-NSEs}
\begin{equation}
M_0=\sum_j f_j^{eq}=\sum_j f_j+\frac{\Delta t \lambda_1}{2} S_G(\mathbf{x}, t),~ \mathbf{M}_1=\sum_j \mathbf{c}_j f_j^{eq}=\sum_j \mathbf{c}_j f_j+\frac{\Delta t \lambda_2}{2} \mathbf{F}_G(\mathbf{x}, t),
\end{equation}
\begin{equation}
\mathbf{M}_2=\sum_j \mathbf{c}_j \mathbf{c}_j f_j^{eq},~ \mathbf{M}_3=\sum_j \mathbf{c}_j \mathbf{c}_j \mathbf{c}_j f_j^{eq},
\end{equation}
\begin{equation}
M_{0F}=\sum_j F_j,~ \mathbf{M}_{1F}=\sum_j \mathbf{c}_j F_j,~ \mathbf{M}_{2F}=\sum_j \mathbf{c}_j \mathbf{c}_j F_j,
\end{equation}
\begin{equation}
M_{0G}=\sum_j G_j,~ \mathbf{M}_{1G}=\sum_j \mathbf{c}_j G_j,~ \mathbf{M}_{2G}=\sum_j \mathbf{c}_j \mathbf{c}_j G_j,
\end{equation}
\begin{equation}
\sum_j\mathbf{e}_j\mathbf{\Lambda}_{jk} = s_0\mathbf{e}_k, ~  \sum_j \mathbf{c}_j\mathbf{\Lambda}_{jk} = \mathbf{S}_{10} \mathbf{e}_k + \mathbf{S}_1 \mathbf{c}_k,
~\sum_j\mathbf{c}_j\mathbf{c}_j\mathbf{\Lambda}_{jk} =\mathbf{S}_{20} \mathbf{e}_k + \mathbf{S}_{21} \mathbf{c}_k+\mathbf{S}_{2}\mathbf{c}_k \mathbf{c}_k,
\end{equation}
\end{subequations}
where $\mathbf{M}_k$, $\mathbf{M}_{kF}$ and $\mathbf{M}_{kG}$ ($k\geq 0$) are the $k$-th moments of $f_j^{eq}, F_j$ and $G_j$, respectively.
 $\mathbf{S}_{10}$ is a $d\times 1$ matrix, $\mathbf{S}_1$ is an invertible $d\times d$ relaxation sub-matrix, $\mathbf{S}_{20}$ and $\mathbf{S}_{21}$ are two $d^{2}\times1$ and $d^{2}\times d$ matrices, and $\mathbf{S}_2$ is an invertible $d^2\times d^2$ relaxation sub-matrix corresponding to the dynamic and bulk viscosities.
 Additionally, Eq. (\ref{eq:M-NSEs}a) gives the following moments of nonequilibrium,
\begin{equation}\label{eq:M-NSEs-1}
M_0^{ne}=\sum_j f_j^{ne}=-\frac{\Delta t \lambda_1}{2} S_{G}(\mathbf{x}, t), ~ \mathbf{M}_1^{ne}=\sum_j \mathbf{c}_j f_j^{ne}=-\frac{\Delta t \lambda_2}{2} \mathbf{F}_{G}(\mathbf{x}, t).
\end{equation}

It should be noted that in the design and analysis of the LB model, moments are usually treated as tensors, which is natural for the SRT-LB model and does not cause any confusion. However, for the MRT-LB model, the introduction of
the collision matrix and the relaxation (sub-) matrices requires converting the velocity tensors which constitute the transformation matrix $\mathbf{M}$ in to vectors with the corresponding dimensions. Only in this way can matrix operations be performed. For the sake of convenience, we introduce the following matrices, which are the first three sub-matrices of the transformation matrix $\mathbf{M}$ \cite{ChaiShi2020, ChaiShi2023} and correspond respectively to the 0th, 1st and 2nd order moments.
\begin{subequations}\label{eq:4-7}
\begin{equation}
\mathbf{e}=(1,1,\cdots,1)=(\mathbf{e}_k)_{1\times q},
\end{equation}
\begin{equation}
\mathbf{E}=(\mathbf{c}_0,\mathbf{c}_1,\cdots,\mathbf{c}_{q-1})=(\mathbf{c}_k)_{d\times q},
\end{equation}
\begin{equation}
\langle \mathbf{EE} \rangle=(\mathbf{c}_0\mathbf{c}_0,\mathbf{c}_1\mathbf{c}_1,\cdots,\mathbf{c}_{q-1}\mathbf{c}_{q-1})=(\mathbf{c}_k\mathbf{c}_k)_{d^2\times q},
\end{equation}
\end{subequations}
where $\mathbf{e}_k$, $\mathbf{c}_k$ and $\mathbf{c}_k\mathbf{c}_k$ are the $k$-th column of $\mathbf{e}$, $\mathbf{E}$ and $\langle \mathbf{EE} \rangle$, respectively. It means that $\mathbf{c}_k$ and $\mathbf{c}_k\mathbf{c}_k$ are taken as $d\times 1$ and $d^{2}\times 1$ vectors. Thus, Eq. (\ref{eq:M-NSEs}e) can be equivalently expressed by matrix operations as
\begin{equation}\label{eq:M-Collision}
\mathbf{e} \mathbf{\Lambda} =s_0\mathbf{e},~\mathbf{E} \mathbf{\Lambda} =\mathbf{S}_{10} \mathbf{e} + \mathbf{S}_1 \mathbf{E},~\langle \mathbf{EE}\rangle \mathbf{\Lambda} =\mathbf{S}_{20} \mathbf{e} + \mathbf{S}_{21} \mathbf{E} + \mathbf{S}_2\langle \mathbf{EE}\rangle.
\end{equation}

Summing Eq. (\ref{eq:4-3}a) and Eq. (\ref{eq:4-5}), and adopting Eqs. (\ref{eq:M-NSEs}) and (\ref{eq:M-NSEs-1}), one can obtain
\begin{subequations}\label{eq:ME-NSEs}
\begin{eqnarray}
\partial_t M_0+\nabla\cdot \mathbf{M}_1 &=& -\frac{s_0}{\Delta t} M_0^{ne}+ M_{0G}+M_{0F}+O(\Delta t) \nonumber\\ &=& \frac{1}{2}s_0 \lambda_1 S_{G}+ M_{0G}+M_{0F}+O(\Delta t),
\end{eqnarray}
\begin{eqnarray}
\partial_t M_0+\nabla\cdot \mathbf{M}_1  +\partial_t(1-s_0/2)M_0^{ne}+\nabla\cdot \big[(\mathbf{I-S}_1 /2)\mathbf{M}_1^{ne} -\mathbf{S}_{10}M_0^{ne}/2\big]\nonumber\\
+ \frac{\Delta t}{2} \partial_t(M_{0G}+(1-\theta)M_{0F}) +  \frac{\Delta t}{2} \nabla\cdot (\mathbf{M}_{1G}+(1-\gamma)\mathbf{M}_{1F})\nonumber\\
=\partial_t M_0+\nabla\cdot \mathbf{M}_1 -\frac{\Delta t}{2}\big[\partial_t(1-s_0/2)\lambda_1 S_G+\nabla\cdot ((\mathbf{I-S}_1 /2)\lambda_2\mathbf{F}_G -\lambda_1\mathbf{S}_{10}S_G/2)\big]\nonumber\\
+ \frac{\Delta t}{2} \partial_t(M_{0G}+(1-\theta)M_{0F}) +  \frac{\Delta t}{2} \nabla\cdot (\mathbf{M}_{1G}+(1-\gamma)\mathbf{M}_{1F})\nonumber\\
= \frac{1}{2}s_0 \lambda_1 S_{G}+ M_{0G}+M_{0F}+O(\Delta t^2).
\end{eqnarray}
\end{subequations}
Taking
\begin{subequations}\label{eq:M-NSEs-2}
\begin{equation}
M_{0G}=(1-\frac{1}{2}s_0\lambda_1)S_G, ~M_{0F}=S_F, ~S=S_G+S_F,
\end{equation}
\begin{equation}
\mathbf{M}_{1G}+\frac{\lambda_1 }{2}\mathbf{S}_{10}S_G=(\mathbf{I}-\frac{1}{2}\mathbf{S}_1\lambda_2)\mathbf{F}_G, ~\mathbf{M}_{1F}=\mathbf{F}_F, ~\mathbf{F}=\mathbf{F}_G+\mathbf{F}_F,
\end{equation}
\end{subequations}
Eq. (\ref{eq:ME-NSEs}) becomes
\begin{subequations}\label{eq:ME-NSEs-1}
\begin{equation}
\partial_t M_0+\nabla\cdot \mathbf{M}_1= S+O(\Delta t),
\end{equation}
\begin{eqnarray}
\partial_t M_0+\nabla\cdot \mathbf{M}_1 +\frac{\Delta t}{2}\big[\partial_t((1-\lambda_1) S_G+(1-\theta) S_{F})\big]\nonumber\\
=S+\nabla\cdot ((1-\lambda_2) \mathbf{F}_G+(1-\gamma)\mathbf{F}_F)\big]+O(\Delta t^2),
\end{eqnarray}
\end{subequations}
which corresponds to the continuity equation in NSEs.

Multiplying $\mathbf{c}_j$ on both sides of Eqs. (\ref{eq:4-3}a) and (\ref{eq:4-5}), and through a summation over $j$, we have
\begin{subequations}\label{eq:ME-NSEs-2}
\begin{eqnarray}
\partial_t \mathbf{M}_1+\nabla\cdot \mathbf{M}_2 = \frac{1}{2}( \lambda_2 \mathbf{S}_1 \mathbf{F}_{G}+\lambda_1 \mathbf{S}_{10}S_G)+ \mathbf{M}_{1G}+\mathbf{M}_{1F}+O(\Delta t)=\mathbf{F}+O(\Delta t),
\end{eqnarray}
\begin{eqnarray}
\partial_t \mathbf{M}_1+\nabla\cdot \mathbf{M}_2 +\partial_t\big[(\mathbf{I-S}_1/2)\mathbf{M}_1^{ne}-\mathbf{S}_{10}M_0^{ne}/2\big]+\nabla\cdot \big[(\mathbf{I-S}_2 /2)\mathbf{M}_2^{ne}-(\mathbf{S}_{21}\mathbf{M}_1^{ne} +\mathbf{S}_{20}M_0^{ne})/2\big]\nonumber\\
+ \frac{\Delta t}{2} \partial_t(\mathbf{M}_{1G}+(1-\theta)\mathbf{M}_{1F}) +  \frac{\Delta t}{2} \nabla\cdot (\mathbf{M}_{2G}+(1-\gamma)\mathbf{M}_{2F})\nonumber\\
=\partial_t \mathbf{M}_1+\nabla\cdot \mathbf{M}_2 -\frac{\Delta t}{2}\big[\partial_t((\mathbf{I-S}_1/2)\lambda_2 \mathbf{F}_G-\mathbf{S}_{10}\lambda_1S_G/2)
-\nabla\cdot (\mathbf{S}_{21}\lambda_2\mathbf{F}_G/2+\mathbf{S}_{20}\lambda_1S_G/2)\big] \nonumber\\
+\nabla\cdot (\mathbf{I-S}_2/2)\mathbf{M}_2^{ne}+ \frac{\Delta t}{2} \partial_t(\mathbf{M}_{1G}+(1-\theta)\mathbf{M}_{1F}) +  \frac{\Delta t}{2} \nabla\cdot (\mathbf{M}_{2G}+(1-\gamma)\mathbf{M}_{2F})\nonumber\\
=\partial_t \mathbf{M}_1+\nabla\cdot \mathbf{M}_2 +\frac{\Delta t}{2}\big[\partial_t((1-\theta) \mathbf{F}_F+(1-\lambda_2) \mathbf{F}_G) \big] \nonumber\\
+\nabla\cdot (\mathbf{I-S}_2/2)\mathbf{M}_2^{ne}+  \frac{\Delta t}{2} \nabla\cdot (\mathbf{M}_{2G}+(1-\gamma)\mathbf{M}_{2F}+(\lambda_1 \mathbf{S}_{20}S_G+\lambda_2 \mathbf{S}_{21}\mathbf{F}_G)/2)\nonumber\\
= \frac{1}{2}( \lambda_2 \mathbf{S}_1 \mathbf{F}_{G}+\lambda_1 \mathbf{S}_{10}S_G)+ \mathbf{M}_{1G}+\mathbf{M}_{1F}+O(\Delta t^2)\nonumber\\
= \mathbf{F}+O(\Delta t^2),
\end{eqnarray}
\end{subequations}
where Eq. (\ref{eq:M-NSEs-2}) is used.

From Eqs. (\ref{eq:4-3}a) and (\ref{eq:M-NSEs}), we have
\begin{equation}\label{eq:M-NSEs-4}
\mathbf{M}_2^{ne}=-\Delta t \mathbf{S}_2^{-1}\big[\partial_t \mathbf{M}_2+\nabla\cdot \mathbf{M}_3-\mathbf{\bar{M}}_{2G}-\mathbf{\mathbf{M}}_{2F}\big]+O(\Delta t^2).
\end{equation}
where
\begin{equation}\label{eq:M-NSEs-4-0}
\mathbf{\bar{M}}_{2G}=\mathbf{M}_{2G}+(\lambda_1 \mathbf{S}_{20}S_G+\lambda_2 \mathbf{S}_{21}\mathbf{F}_G)/2.
\end{equation}

Substituting Eq. (\ref{eq:M-NSEs-4}) into Eq. (\ref{eq:ME-NSEs-2}b), we can obtain the moment equation corresponding to the momentum equation in NSEs
\begin{eqnarray}\label{eq:ME-NSEs-3}
\partial_t \mathbf{M}_1+\nabla\cdot \mathbf{M}_2 +\frac{\Delta t}{2}\partial_t[(1-\theta) \mathbf{F}_F+(1-\lambda_2) \mathbf{F}_G)] \nonumber\\
= \mathbf{F} +\Delta t \nabla\cdot \big[(\mathbf{S}_2^{-1}-\mathbf{I}/2)(\partial_t \mathbf{M}_2+\nabla\cdot\mathbf{M}_3)-\mathbf{S}_2^{-1}(\mathbf{\bar{M}}_{2G}+\mathbf{M}_{2F})+\gamma\mathbf{M}_{2F}\big]+O(\Delta t^2).
\end{eqnarray}

Let
\begin{equation}\label{eq:M-NSEs-5}
(1-\lambda_1)S_G+(1-\theta)S_F=0,(1-\lambda_2)\mathbf{F}_G+(1-\gamma)\mathbf{F}_F=0,(1-\lambda_2)\mathbf{F}_G+(1-\theta)\mathbf{F}_F=0,\mathbf{M}_{2F}=0,
\end{equation}
Eqs. (\ref{eq:ME-NSEs-1}b) and (\ref{eq:ME-NSEs-3}) become
\begin{subequations}\label{eq:ME-NSEs-4}
\begin{equation}
\partial_t M_0+\nabla\cdot \mathbf{M}_1= S+O(\Delta t^2),
\end{equation}
\begin{eqnarray}\label{eq:ME-NSEs-4b}
\partial_t \mathbf{M}_1+\nabla\cdot \mathbf{M}_2 = \mathbf{F} +\Delta t \nabla\cdot \big[(\mathbf{S}_2^{-1}-\mathbf{I}/2)(\partial_t \mathbf{M}_2+\nabla\cdot\mathbf{M}_3)-\mathbf{S}_2^{-1}\mathbf{\bar{M}}_{2G}\big]+O(\Delta t^2),
\end{eqnarray}
\end{subequations}
where the relaxation sub-matrix $\mathbf{S}_2$ and auxiliary source term $\mathbf{\bar{M}}_{2G}$ need to be determined.

\subsection{The recovery of general Navier-Stokes equations from the moment equations}

Note that the moment equations (\ref{eq:ME-NSEs-4}) are the general forms recovered from MRT-LB method (\ref{eq:2-1}) with the basic moments (\ref{eq:M-NSEs}). Any NSEs recovered from RMRT-LB model must be the special cases of these moment equations. Now, we consider the following $d$-dimensional target NSEs with a general form \cite{Yuan2020}
\begin{subequations}\label{eq:NSE0}
\begin{equation}
\partial_t \bar{\rho}+\nabla\cdot(\rho \mathbf{u})= S,
\end{equation}
\begin{equation}
\partial_t (\rho\mathbf{u})+\nabla\cdot(\rho \mathbf{uu}+p\mathbf{I})= \nabla\cdot \mathbf{\sigma}+ \mathbf{F},
\end{equation}
\end{subequations}
where the viscous shear stress $\mathbf{\sigma}$ is defined by
\begin{eqnarray}\label{eq:sigma0}
\mathbf{\sigma}&=&\mu\big[\nabla \mathbf{u}+(\nabla \mathbf{u})^T\big]+\lambda (\nabla\cdot \mathbf{u})\mathbf{I}\nonumber\\
&=&\mu\big[\nabla \mathbf{u}+(\nabla \mathbf{u})^T-\frac{2}{d}(\nabla\cdot \mathbf{u})\mathbf{I}\big]+\mu_{b} (\nabla\cdot \mathbf{u})\mathbf{I},
\end{eqnarray}
where $\mu$ is the dynamic viscosity, $\lambda=\mu_{b}-2\mu/d$ with $\mu_{b}$ being the bulk viscosity  \cite{Dellar2001,Kundu2016}. $\bar{\rho}$ is a physical quantity related to $\rho$ or a constant.

Eq. (\ref{eq:NSE0}) contains the common NSEs for governing incompressible and compressible flows in both single-phase and multiphase systems \cite{Yuan2020}. In order to obtain the NSEs (\ref{eq:NSE0}) from the moment equations (\ref{eq:ME-NSEs-4}), let
\begin{subequations}\label{eq:M-NSEs-6a}
\begin{equation}
\theta=\gamma=\lambda_1=\lambda_2=1,
\end{equation}
\begin{equation}
M_0=\bar{\rho},\quad \mathbf{M}_1=\rho\mathbf{u},\quad \mathbf{M}_2= \rho\mathbf{uu}+p\mathbf{I},
\end{equation}
\end{subequations}
Eq. (\ref{eq:ME-NSEs-4}) becomes
\begin{subequations}\label{eq:eq:NSEs0-1}
\begin{equation}
\partial_t \bar{\rho}+\nabla\cdot(\rho \mathbf{u})= S + O(\Delta t^2),
\end{equation}
\begin{equation}
\partial_t (\rho\mathbf{u})+\nabla\cdot(\rho \mathbf{uu}+p\mathbf{I})= \mathbf{F} +\Delta t \nabla\cdot \big[(\mathbf{S}_2^{-1}-\mathbf{I}/2)(\partial_t (\rho \mathbf{uu}+p\mathbf{I})+\nabla\cdot\mathbf{M}_3)-\mathbf{S}_2^{-1}\mathbf{\bar{M}}_{2G}\big] + O(\Delta t^2),
\end{equation}
\end{subequations}
where $\mathbf{M}_3$ and $\mathbf{\bar{M}}_{2G}$ have different expressions for incompressible and compressible cases.

$\mathbf{M}_3$ is usually composed of two kinds of moments: viscosity-related moment $\mathbf{M}_{3v}$ and auxiliary moment $\mathbf{M}_{30}$. Auxiliary moments $\mathbf{M}_{30}$ and $\mathbf{\bar{M}}_{2G}$ (or $\mathbf{M}_{2G}$ (\ref{eq:M-NSEs-4-0})) are used to eliminate spurious terms or to recover the desired terms. The decomposed form of $\mathbf{M}_3$ represents two modeling methods: directly handling anisotropy in $\mathbf{M}_{3v}$ or placing the term(s) causing anisotropy in the auxiliary moment $\mathbf{M}_{30}$. The former is processed directly with the relaxation matrix $\mathbf{S}_2$, while the latter involves some gradients and requires approximate calculation.

For incompressible fluids,
\begin{subequations}\label{eq:M-NSEs-6}
\begin{equation}
\mathbf{M}_3=\rho(c_s^2 \mathbf{\mathbf{\Delta}}+k\bar{\delta}^{(4)}) \cdot \mathbf{u} +\mathbf{M}_{30},
\end{equation}
\begin{equation}
\mathbf{\bar{M}}_{2G}=(\mathbf{I}-\mathbf{S}_2/2)\big[\partial_t (p-\bar{\rho} c_s^2)\mathbf{I}+\partial_t (\rho \mathbf{uu})+ c_s^2(\mathbf{u} \nabla\rho +(\mathbf{u} \nabla\rho)^{T})+(k\bar{\delta}^{(4)}\cdot \mathbf{u})\cdot\nabla \rho+c_s^2 S\mathbf {I}+\nabla\cdot\mathbf{M}_{30})\big],
\end{equation}
\end{subequations}
where $c_s$ represents the sound speed, $\Delta_{\alpha\beta\gamma\theta}=\delta_{\alpha\beta}\delta_{\gamma\theta}+\delta_{\alpha\gamma}\delta_{\beta\theta}+\delta_{\beta\gamma}\delta_{\alpha\theta}$, $k$ is a parameter with $k=0$ or $1$, corresponding to isotropy or anisotropy. $\mathbf{M}_{30}$ is an auxiliary 3rd moment. $\mathbf{\bar{\delta}}^{(4)}$ is caused by the anisotropy of the lattice tensor, and is given by \cite{ChaiShi2023}
\begin{equation}
\mathbf{\bar{\delta}}^{(4)}_{\alpha\beta\gamma\theta}=c_\alpha^2-3c_{s}^2,~\alpha=\beta=\gamma=\theta;\\
\mathbf{\bar{\delta}}^{(4)}_{\alpha\beta\gamma\theta}=0, ~else,
\end{equation}
where
$c_\alpha=\Delta x_{\alpha}/\Delta t$ ($\alpha=1, 2, \ldots, d$) in $d$-dimensional space with $\Delta x_{\alpha}$ being the spacing step in $\alpha$ axis. There are two ways to compute the pressure $p$ : set $p=\rho c_s^2$ for weak compressible fluids or compute $p$ as an independent variable for incompressible fluids.

For compressible fluids, the equation of state can be expressed as $p=\rho RT$, then $\mathbf{M}_3$ and $\mathbf{\bar{M}}_{2G}$ can be rewritten as
\begin{subequations}\label{eq:M-NSEs-6-1}
\begin{equation}
\mathbf{M}_3=p( \mathbf{\mathbf{\Delta}}+k\tilde{\delta}^{(4)}) \cdot \mathbf{u} +\mathbf{M}_{30},
\end{equation}
\begin{equation}
\mathbf{\bar{M}}_{2G}=(\mathbf{I}-\mathbf{S}_2/2)\big[\partial_t (p\mathbf{I}+\rho \mathbf{uu})+(\mathbf{u} \nabla p +(\mathbf{u} \nabla p)^{T})+(k\tilde{\delta}^{(4)}\cdot \mathbf{u})\cdot\nabla p+\nabla\cdot(p\mathbf {uI}+\mathbf{M}_{30})\big],
\end{equation}
\end{subequations}
where $\tilde{\delta}^{(4)}=\bar{\delta}^{(4)}/{c_s^2}$.

Finally, we can obtain the following NSEs,
\begin{subequations}\label{eq:NSE}
\begin{equation}
\partial_t \bar{\rho}+\nabla\cdot(\rho \mathbf{u})= S+O(\Delta t^2),
\end{equation}
\begin{equation}
\partial_t (\rho\mathbf{u})+\nabla\cdot(\rho \mathbf{uu}+p\mathbf{I})= \nabla\cdot \mathbf{\sigma}+ \mathbf{F}+O(\Delta t^2),
\end{equation}
\end{subequations}
with $\mathbf{\sigma}$ satisfying
\begin{subequations}\label{eq:sigma}
\begin{equation}
\mathbf{\sigma}=\Delta t c_s^2 \rho \big(\mathbf{S}_2^{-1}-\mathbf{I}/2\big) \big[\big(\nabla\mathbf{u}+(\nabla\mathbf{u})^{T}\big)+\nabla\cdot(k \bar{\delta}^{(4)}\cdot \mathbf{u})\big], ~ for ~ incompressible ~fluids;
\end{equation}
\begin{equation}
\mathbf{\sigma}=\Delta t p \big(\mathbf{S}_2^{-1}-\mathbf{I}/2\big) \big[\big(\nabla\mathbf{u}+(\nabla\mathbf{u})^{T}\big)+\nabla\cdot(k \tilde{\delta}^{(4)}\cdot \mathbf{u})\big], ~ for ~compressible ~fluids,
\end{equation}
\end{subequations}
which needs to be determined by selecting proper relaxation matrix $\mathbf S_2^{-1}$ \cite{ChaiShi2023}.

Let
\begin{equation}\label{eq:4-27}
\mathbf{S}_2^{-1}=
\left(
\begin{array}{cc}
    \mathbf{S}_2^{(1)} & 0  \\
    0 &  \mathbf{S}_2^{(2)}  \\
\end{array}
\right),
\end{equation}
with
\begin{equation}\label{eq:4-27-1}
\mathbf{S}_2^{(1)}=\mathbf{diag} (s_{s\alpha}^{-1})+\mathbf{a}\mathbf{b}^{T}/d, \mathbf{S}_2^{(2)}=\mathbf{diag}(s_{\alpha\beta}^{-1})_{\alpha \neq \beta},
\end{equation}
where $\mathbf{a}=(a_{\alpha}), \mathbf{b}=(b_{\beta})$ with $a_{\alpha}=(s_{b\alpha }^{-1}-s_{s\alpha}^{-1})(c_{\alpha}^2-c_{s}^2)$ and $b_{\beta}=1/(c_{\beta}^2-c_{s}^2)$, then substituting Eq. (\ref{eq:4-27}) into Eq. (\ref{eq:sigma}), one can obtain the dynamic and bulk viscosities (${\mu}$ and ${\mu_b}$)
\begin{subequations}\label{eq:4-29}
\begin{eqnarray}
\mu= \big(s_{\alpha \beta}^{-1}-\frac{1}{2}\big) \rho c_s^2\Delta t,\alpha\neq \beta,\ \ \mu= \frac{1}{2}\big(s_{s\alpha}^{-1}-\frac{1}{2}\big) \rho\left[ k c_\alpha^2+ (2-3k)c_s^2\right]\Delta t,\nonumber\\
\mu_{b}= \frac{1}{d}\big(s_{b\alpha }^{-1}-\frac{1}{2}\big) \rho\left[ k c_\alpha^2+ (2-3k)c_s^2\right]\Delta t, ~ for ~ incompressible ~fluids;
\end{eqnarray}
\begin{eqnarray}
\mu= \big(s_{\alpha \beta}^{-1}-\frac{1}{2}\big) p\Delta t,\alpha\neq \beta,\ \ \mu= \frac{1}{2}\big(s_{s\alpha}^{-1}-\frac{1}{2}\big) p\left[ k c_\alpha^2/c_s^2+ (2-3k)\right]\Delta t,\nonumber\\
\mu_{b}= \frac{1}{d}\big(s_{b\alpha }^{-1}-\frac{1}{2}\big) p\left[ k c_\alpha^2/c_s^2+ (2-3k)\right]\Delta t, ~ for ~ compressible ~fluids.
\end{eqnarray}
\end{subequations}
Thus, the NSEs (\ref{eq:NSE0}) is recovered in order $O(\Delta t^2)$.

Note that the auxiliary moment $\mathbf{\bar{M}}_{2G}$ in Eq. (\ref{eq:M-NSEs-6}b) or Eq. (\ref{eq:M-NSEs-6-1}b) is complete and does not ignore any terms. It can be further simplified by approximating $\partial_t (\rho \mathbf{uu})$, or using other conditions.

For incompressible fluids, with the help of $\partial_t (\rho \mathbf{uu})$ which is given by
\begin{equation}\label{eq:rhoUU}
\partial_t (\rho \mathbf{uu})=\mathbf{u}\bar{\mathbf F}+\bar{\mathbf F}\mathbf{u}-c_s^2 [\mathbf{u}\nabla \rho+(\mathbf{u}\nabla \rho)^T]-\nabla \cdot(\rho\mathbf{uuu})-\mathbf{uu}\bar S + O(\Delta t \mathbf{u}),
\end{equation}
where $\bar{\mathbf F}=\mathbf F+\nabla(\rho c_s^2-p)$, $\bar S=S+\partial_t(\rho-\bar{\rho})$, then $\mathbf{\bar{M}}_{2G}$ can be rewritten as
\begin{equation}\label{eq:M2G}
\mathbf{\bar{M}}_{2G}=(\mathbf{I}-\mathbf{S}_2/2)\big[\partial_t (p-\bar{\rho} c_s^2)\mathbf{I}+ (k\bar{\delta}^{(4)}\cdot \mathbf{u})\cdot\nabla \rho+c_s^2 S\mathbf {I}+\mathbf{u}\bar{\mathbf F}+\bar{\mathbf F}\mathbf{u}-\mathbf{uu}\bar S+\nabla\cdot(\mathbf{M}_{30}-\rho\mathbf{uuu})\big],
\end{equation}
where $O(\Delta t \mathbf{u})$ is omitted in $\mathbf{\bar{M}}_{2G}$.

For compressible fluids, $\partial_t (\rho \mathbf{uu})$ has the following form
\begin{equation}\label{eq:rhoUU-1}
\partial_t (\rho \mathbf{uu})=\mathbf{u}\mathbf F+\mathbf F\mathbf{u}-[\mathbf{u}\nabla p+(\mathbf{u}\nabla p)^T]-\nabla \cdot(\rho\mathbf{uuu})-\mathbf{uu}\bar S + O(\Delta t \mathbf{u}),
\end{equation}
and $\mathbf{\bar{M}}_{2G}$ becomes
\begin{equation}\label{eq:M2G-1}
\mathbf{\bar{M}}_{2G}=(\mathbf{I}-\mathbf{S}_2/2)\big[\partial_t p\mathbf{I}+ (k\tilde{\delta}^{(4)}\cdot \mathbf{u})\cdot\nabla \rho +\mathbf{u}\mathbf F+\mathbf F\mathbf{u}-\mathbf{uu}\bar S+\nabla\cdot(p\mathbf{uI}+\mathbf{M}_{30}-\rho\mathbf{uuu})\big],
\end{equation}
where $O(\Delta t \mathbf{u})$ is also omitted.

\subsection{Some special cases of RMRT-LB model for incompressible Navier-Stokes equations}

The present RMRT-LB model is a unified one that incorporates incompressible and compressible cases. It contains several existing LB models and generalizes them. In this section, we consider the incompressible case.

For the standard lattice models (or single-layer velocity models), the present model generalizes the model of Ref. \cite{Yuan2020} from a square, isotropic SRT-LB model to a rectangular, anisotropic MRT-LB model. It also extends the model by Chai et al. \cite{ChaiShi2020,ChaiShi2023}, which employs a standard lattice model to solve the weakly compressible NSEs [$k=1$ in Eq. (\ref{eq:M-NSEs-6})]. Let $\mathbf M_{30}=\mathbf 0$, $p=\rho c_s^2$, and neglect the terms of $O(Ma^3)$ in Eq. (\ref{eq:M2G}), then $\mathbf{\bar{M}}_{2G}$ can be written as
 \begin{equation}\label{eq:M2Ga}
\mathbf{\bar{M}}_{2G}=(\mathbf{I}-\mathbf{S}_2/2)\big[\partial_t (p-\bar{\rho} c_s^2)\mathbf{I}+ (\bar{\delta}^{(4)}\cdot \mathbf{u})\cdot\nabla \rho+c_s^2 S\mathbf {I}+\mathbf{u}\bar{\mathbf F}+\bar{\mathbf F}\mathbf{u}-\mathbf{uu}\bar S\big],
 \end{equation}
 which is simplified by taking $\bar{\rho}=\rho$ as
 \begin{equation}\label{eq:M2Ga-1}
\mathbf{\bar{M}}_{2G}=(\mathbf{I}-\mathbf{S}_2/2)\big[ \mathbf{u}\mathbf F+\mathbf F\mathbf{u}+(c_s^2 \mathbf {I}-\mathbf{uu}) S\big],
 \end{equation}
where $(\bar{\delta}^{(4)}\cdot \mathbf{u})\cdot\nabla \rho=O(Ma^3)$ is used.

 \emph{Remark 1.} As did in Peng et al.'s model \cite{PengGuoWang2019}, another way to treat $\mathbf M_3$ is to put $\bar{\delta}^{(4)}$ in Eq. (\ref{eq:M-NSEs-6}) into $\mathbf M_{30}$, which means $k=0$ in $\mathbf M_{3}$ and $\mathbf M_{30}=\rho \bar{\delta}^{(4)} \cdot \mathbf{u}$. This results in some gradients in $\mathbf {\bar M}_{2G}$, which requires additional calculations. Note that if $\mathbf S_2$ is properly used as in Eq. (\ref{eq:4-27}), $\mathbf {\bar M}_{2G}$ is much simpler (\ref{eq:M2Ga-1}).

 Taking $\bar{\rho}=\rho=const$ (e.g., $\bar{\rho}=\rho=1$) and $S=0$ in Eq. (\ref{eq:NSE}), one can obtain the following incompressible NSEs
\begin{subequations}\label{eq:NSE2}
\begin{equation}
\nabla\cdot \mathbf{u}= 0,
\end{equation}
\begin{equation}
\partial_t \mathbf{u}+\nabla\cdot(\mathbf{uu}+p\mathbf{I})= \nabla\cdot [\nu(\nabla \mathbf{u}+(\nabla \mathbf{u})^T)]+ \mathbf{F}.
\end{equation}
\end{subequations}
For classical incompressible multiphase flow systems, the flow field can be described as \cite{Yuan2020}
\begin{subequations}\label{eq:NSE3}
\begin{equation}
\nabla\cdot \mathbf{u}= 0,
\end{equation}
\begin{equation}
\partial_t (\rho\mathbf{u})+\nabla\cdot(\rho \mathbf{uu}+p\mathbf{I})= \nabla\cdot \sigma+ \mathbf{F}.
\end{equation}
\end{subequations}
Eq. (\ref{eq:NSE3}) can also be derived from Eq. (\ref{eq:NSE}), if $S=\mathbf u \cdot \nabla \rho$ and $\bar \rho=const$.

In addition, let $S=0, ~\mathbf F=-\rho \nabla \mu$, and $\mathbf M_2=\rho\mathbf{uu}$, the present model can also derive the macroscopic equations
\begin{subequations}
\begin{equation}
\partial_t {\rho}+\nabla\cdot(\rho \mathbf{u})= 0,
\end{equation}
\begin{equation}
\partial_t (\rho\mathbf{u})+\nabla\cdot(\rho \mathbf{uu})=-\rho \nabla \mu+ \nabla\cdot \bar{ \mathbf{\sigma}},
\end{equation}
\end{subequations}
in Ref. \cite{Guo2021}, where the viscous stress is $\bar{\sigma}_{\alpha\beta}=\rho \nu\left[\partial_{\alpha}u_{\beta}+\partial_{\beta}u_{\alpha}+(\nabla \cdot \mathbf u)\delta_{\alpha\beta} \right]$. We would like to point out that the model in Ref. \cite{Guo2021} is a SRT version and does not yield an accurate bulk viscosity.

The current model also extends the models by Chai et al. \cite{ChaiShi2020,ChaiShi2023}, which employs a standard velocity model to solve the weakly compressible NSEs. The governing equations of the weakly compressible NSEs can be obtained simply by letting $\bar{\rho}=\rho$, $\mathbf M_{30}=\mathbf 0$ and $p=\rho c_s^2$, which can be written as
\begin{subequations}\label{eq:NSE1}
\begin{equation}
\partial_t {\rho}+\nabla\cdot(\rho \mathbf{u})= S,
\end{equation}
\begin{equation}
\partial_t (\rho\mathbf{u})+\nabla\cdot(\rho \mathbf{uu}+p\mathbf{I})= \nabla\cdot \mathbf{\sigma}+ \mathbf{F}.
\end{equation}
\end{subequations}
In addition, it can be found that $\mathbf{\bar{M}}_{2G}$ [Eq. (\ref{eq:M2Ga-1})] is the same expression as in Ref. \cite{ChaiShi2023}.

 For the multispeed lattice models in the incompressible case, the current model is still applicable ($k=0$, $\mathbf M_{30}=\rho\mathbf{uuu}$ and $\mathbf M_{30}$ can be dropped in this case), and the isotropy condition can hold. Although a multispeed lattice model can be used for incompressible fluids, it is better to use a standard lattice from the point of view of reducing computational cost.



\subsection{Some special cases of RMRT-LB model for compressible Navier-Stokes equations}

In the classic LB model (such as D$2$Q$9$ or D$3$Q$27$ lattice model), the discrete velocities have only one value in each direction and lack constraints on the energy, so they can usually only simulate isothermal, low-velocity, weakly compressible flows. Next, we consider the compressible case. For the standard lattice models, some researchers \cite{Prasianakis2007, Li2012} have dealt with compressible flows by placing certain terms in the auxiliary source term $\mathbf M_{2G}$ into the equilibrium. In addition, these models usually adopt square lattice and are SRT versions. By contrast, the present model is a rectangular MRT version and uses $\mathbf M_{2G}$ to handle the higher-order terms of Mach numbers.

The multispeed lattice model is mainly used to simulate compressible flows. Based on introducing correction terms in the kinetic equations, one can eliminate the spurious terms in the momentum equation resulting from the constraints of the standard lattices. In this case, $k=0$, $\mathbf M_{30}=\rho\mathbf{uuu}$ and $p=\rho RT$ is required, and $\partial_t (\rho \mathbf{uu})$ can be written as Eq. (\ref{eq:rhoUU-1}), and Eq. (\ref{eq:M2G-1}) becomes
\begin{equation}
\mathbf{\bar{M}}_{2G}=(\mathbf{I}-\mathbf{S}_2/2)\big[\partial_t p\mathbf{I}+\mathbf{u}{\mathbf F}+{\mathbf F}\mathbf{u}-\mathbf{uu}\bar S+\nabla\cdot(p\mathbf {uI})\big].
\end{equation}
We would like to point out that the present model can derive the correct moment equations and macroscopic equations.

For the standard lattice model, $\mathbf{\tilde{\delta}}^{(4)}$ in Eq. (\ref{eq:M-NSEs-6-1}) can be absorbed into $\mathbf{M}_{30}$ \cite{Prasianakis2007, Li2012}, that is, $k=0$, $\mathbf M_{30}=p \tilde{\delta}^{(4)} \cdot \mathbf{u}$. From this, the SRT/MRT-LB model applicable to compressible flows with a standard lattice model can be derived.


Next, we focus on the multi-layer velocity model in which multispeed (higher-order) lattices are adopted in order to adequately represent all the moments pertinent to the recovery of the full NSEs. This model is usually used for simulating compressible flows. In the common D$2$Q$9$ lattice model, the third-order moments are incomplete. To make the third-order moments complete, a multispeed lattice can be used, such as D$2$Q$17$ lattice model.  A third order EDF is needed in the multi-layer velocity model, and the construction of the third order EDF is discussed in the next subsection.


\section{The equilibrium, auxiliary and source distribution functions of RMRT-LB method}

From above analysis, one can clearly observe that to recover the macroscopic NSEs (\ref{eq:NSE0}), the equilibrium, auxiliary and source distribution functions should satisfy some necessary requirements.  Once the zeroth- to third-order moments of the equilibrium,  the zeroth- to second-order moments of $G_j$, and the zeroth- to first-order moments of $F_j$ are specified, the corresponding equilibrium distribution function can be obtained. 
 Here $\mathbf{M} \in R^{b\times b} $ is an invertible transformation matrix related to the collision matrix $\mathbf {\Lambda}$ ($\mathbf {\Lambda}=\mathbf M^{-1}\mathbf S \mathbf M$), whose rows are composed of discrete velocities in $V_b$. The different structures of collision matrices have been discussed in detail by Chai et al. \cite{ChaiShi2023} and will not be repeated here. $\mathbf{S}$ is a block-lower-triangle matrix which can be written as
 \begin{equation}
\mathbf{S}=(\mathbf{S}_{kj}),\ \ \mathbf{S}_{kj}=0\ (k<j),\ \ \mathbf{S}_{kk}=\mathbf{S}_{k},
\end{equation}
 where $\mathbf{S}_k \in R^{n_k\times n_k}$ is a relaxation matrix corresponding to the $k$-th ($0\leq k\leq m$) order moment of discrete velocity.

The representation of the distribution function on the discrete velocity set (lattice model) is based on the Hermite matrix, rather than the Hermite polynomials. Let $\mathbf H \in R^{b\times b}$ be a Hermite matrix with $\mathbf {H}=(\mathbf{H}_0^T,\mathbf{H}_1^T,\ldots,\mathbf{H}_m^T)^T$. $\mathbf W=\text{diag}\{\omega_k, 0\leq k \leq b\}$ is a weight matrix, then $\mathbf{HWH}^T$ is block-weighted orthogonal which is given by $\mathbf{HWH}^T=\text{diag}\{\mathbf{H}_k \mathbf W \mathbf{H}^T_k, 0\leq k \leq m\}$. Therefore, we have $\mathbf{H}^{-1}=\mathbf{WH}^T(\mathbf{HWH}^T)^{-1}$. When $\mathbf H$ and $\mathbf W$ are given, we can obtain the representation of the required distribution functions, such as the EDF and SDF.

Based on the previous work  \cite{Wu2024}, the general form of EDF or SDF in this work can be written as follows,
\begin{equation}\label{eq:geq}
\mathbf g=\mathbf{H}^{-1}\mathbf{H}\mathbf g=\mathbf{H}^{-1}\mathbf{m}_{H}^{g}=\sum_{k=0}^{m} (\mathbf {W}\mathbf{H}_k^T)(\mathbf{H}_k \mathbf {W}\mathbf{H}_k^T)^{-1} \mathbf {m}_{H,k}^{g}.
\end{equation}

Taking $\mathbf g$ as an EDF vector $\mathbf{f}^{eq}$ or a NEDF vector $\mathbf{f}^{ne}$, one can obtain
\begin{equation}\label{eq:feq}
\mathbf{f}^{eq}=\mathbf{H}^{-1}\mathbf{H}\mathbf{f}^{eq}=\mathbf{H}^{-1}\mathbf{m}_{H}^{eq}=\sum_{k=0}^{m} (\mathbf {W}\mathbf{H}_k^T)(\mathbf{H}_k \mathbf {W}\mathbf{H}_k^T)^{-1} \mathbf {m}_{H,k}^{eq},
\end{equation}
or
\begin{equation}\label{eq:feq-1}
f_i^{eq}=\omega_i \sum_{k=0}^{m} (\mathbf{H}_k^T)_i(\mathbf{H}_k \mathbf {W}\mathbf{H}_k^T)^{-1} \mathbf {m}_{H,k}^{eq},
\end{equation}
and
\begin{equation}\label{eq:Fne}
\mathbf{f}^{ne}=\mathbf{H}^{-1}\mathbf{H}\mathbf{f}^{ne}=\mathbf{H}^{-1}\mathbf{m}_{H}^{ne}=\mathbf {W}\sum_{k=0}^{m} (\mathbf{H}_k^T)(\mathbf{H}_k \mathbf {W}\mathbf{H}_k^T)^{-1} \mathbf {m}_{H,k}^{ne}.
\end{equation}
 Furthermore, based on Eq. (\ref{eq:Fne}), the SRT-collision operator $-(1/{\tau})\mathbf{f}^{ne}$ can be extended to the MRT-collision operator
 \begin{equation}\label{eq:ShanMRT-LB}
-\mathbf {W}\sum_{k=0}^{m} (1/{\tau_k})(\mathbf{H}_k^T)(\mathbf{H}_k \mathbf {W}\mathbf{H}_k^T)^{-1} \mathbf {m}_{H,k}^{ne},
\end{equation}
which can be regarded as the discrete version of Shan et al.'s MRT-collision operator \cite{Shan2007}. It should be pointed out that the block triple-relaxation-time LB (B-TriRT-LB) model \cite{ZhaoShi2020} (which includes the regularized LB model and modified LB model), and even the model defined by Eq. (\ref{eq:ShanMRT-LB}) are all the special cases of the Hermite-moment based MRT-LB (HMRT-LB) model \cite{Wu2024}. In fact, the collision operator of HMRT-LB model can be expressed as
\begin{equation}\label{eq:H-MRT-LB}
-\mathbf{H^{-1}SHf}^{ne}=-s_0\mathbf f^{ne}-\mathbf{H^{-1}}(\mathbf S-s_0\mathbf I)\mathbf{Hf}^{ne}.
\end{equation}
When $\mathbf S$ is block-diagonal, Eq. (\ref{eq:H-MRT-LB}) becomes
\begin{equation}\label{eq:H-MRT-LB-1}
-\mathbf{H^{-1}SHf}^{ne}=-s_0\mathbf f^{ne}-\mathbf {W}\sum_{k=0}^{m} (\mathbf{H}_k^T)(\mathbf{H}_k \mathbf {W}\mathbf{H}_k^T)^{-1} (\mathbf S_k-s_0\mathbf I_k)\mathbf {m}_{H,k}^{ne},
\end{equation}
from which the B-TriRT-LB model can be obtained with $m=2$, or by setting $\mathbf S_k=s_0\mathbf I_k, k > 2$.

\section{Multi-layer velocity lattice models of RMRT-LB method on rectangular lattice}

Defining the $k$th order moment of the velocity set \{$\mathbf{c}_j, ~0 \leq j\leq b$\}=\{$\mathbf 0, \mathbf{c}_j, ~1 \leq j\leq b$\} as
\begin{equation}\label{eq:new0}
\Delta^{(0)}=\sum_{j\geq 0}\omega_j=1,~\Delta_{i_1 i_2 \cdots i_k}^{(k)}=\sum_{j \geq 0}\omega_j \mathbf{c}_{ji_1}\mathbf{c}_{ji_2}\cdots \mathbf{c}_{ji_k}, k \geq 1.
\end{equation}

Due to the symmetry of the lattice model, $\mathbf{c}_{j}$ and $-\mathbf{c}_{j}$ must be included in the velocity set at the same time and their weight coefficients are equal, which leads to odd-order moments always equal to 0.

Consider the following Hermite matrices $\mathbf{H}_0=\mathbf{e}$, $\mathbf{H}_1=\{\mathbf{c}_j\}=\mathbf E$, $\mathbf{H}_2=\{\mathbf c_j \mathbf c_j\} -\mathbf \Delta^{(2)}=\langle \mathbf{EE} \rangle-\mathbf \Delta^{(2)}$, and $\mathbf{H}_3=\{\mathbf c_j \mathbf c_j \mathbf c_j\} -<\mathbf c_j \mathbf \Delta^{(2)}>$,
where $<\mathbf c_j \mathbf \Delta^{(2)}>_{\alpha\beta\gamma}=\mathbf c_{j\alpha} \mathbf \Delta^{(2)}_{\beta\gamma}+\mathbf c_{j\beta} \mathbf \Delta^{(2)}_{\alpha\gamma}+\mathbf c_{j\gamma} \mathbf \Delta^{(2)}_{\alpha\beta}$. Note that due to symmetry, there are multiple identical row vectors in $\mathbf H_k$ for $k>1$, and only one of them is retained. For instance, the row vectors of $\mathbf H_2$, $(H_2)_{\alpha\beta}=\{\mathbf c_{j\alpha} \mathbf c_{j\beta} -\mathbf \Delta^{(2)}_{\alpha\beta}\}=(H_2)_{\beta\alpha}$.
\{$\mathbf{H}_0$, $\mathbf{H}_1$\}, \{$\mathbf{H}_0$, $\mathbf{H}_3$\}, \{$\mathbf{H}_1$, $\mathbf{H}_2$\}, \{$\mathbf{H}_2$, $\mathbf{H}_3$\} are weighted orthogonal due to the symmetry of the lattice model. \{$\mathbf{H}_0$, $\mathbf{H}_2$\} is also weighted orthogonal, then we have \{$\mathbf{H}_0$, $\mathbf{H}_1$, $\mathbf{H}_2$\} is block-weighted orthogonal.

Let $\Delta_{\alpha\alpha}^{(2)}=c_{s\alpha}^2>0$, $\Delta_{\alpha\beta}^{(2)}=0$ ($\alpha \neq \beta$), then internal orthogonality of $\mathbf H_1$ can be written as
\begin{equation}\label{eq:new5}
(\mathbf{H}_1)_{\alpha} \mathbf W(\mathbf{H}_1^{T})_{\beta}=\sum \omega_j \mathbf{c}_{j\alpha} \mathbf{c}_{j\beta}=\Delta_{\alpha\beta}^{(2)}=0, \quad \alpha \neq \beta.
\end{equation}

To satisfy the internal weighted orthogonality of $\mathbf H_2$, the following equation
\begin{equation}\label{eq:new6}
(\mathbf{H}_2)_{\alpha\alpha} \mathbf W (\mathbf{H}_2^{T})_{\beta\beta}=\Delta_{\alpha\alpha\beta\beta}^{(4)}-\Delta_{\alpha\alpha}^{(2)}\Delta_{\beta\beta}^{(2)}=0, \quad \alpha \neq \beta,
\end{equation}
or
\begin{equation}\label{eq:new7}
\Delta_{\alpha\alpha\beta\beta}^{(4)}=\Delta_{\alpha\alpha}^{(2)}\Delta_{\beta\beta}^{(2)}, \quad \alpha \neq \beta,
\end{equation}
should hold.
Further, to make $\mathbf{H}_3$ is block-weighted orthogonal to \{$\mathbf{H}_0$, $\mathbf{H}_1$, $\mathbf{H}_2$\}, we have
\begin{equation}\label{eq:new8}
(\mathbf H_3)_{\alpha\beta\gamma} \mathbf W (\mathbf H_1^{T})_{\theta}=\mathbf \Delta^{(4)}_{\alpha\beta\gamma\theta}-<\mathbf \Delta^{(2)}\mathbf \Delta^{(2)}>_{\alpha\beta\gamma\theta}=0,
\end{equation}
where $<\mathbf \Delta^{(2)} \mathbf \Delta^{(2)}>_{\alpha\beta\gamma\theta}=\mathbf \Delta^{(2)}_{\alpha\beta} \mathbf \Delta^{(2)}_{\gamma\theta}+\mathbf \Delta^{(2)}_{\alpha\gamma} \mathbf \Delta^{(2)}_{\beta\theta}+\mathbf \Delta^{(2)}_{\beta\gamma} \mathbf \Delta^{(2)}_{\alpha\theta}$. Then $\Delta_{\alpha\alpha\alpha\alpha}^{(4)}=3\Delta_{\alpha\alpha}^{(2)}\Delta_{\alpha\alpha}^{(2)}$, and Eq. (\ref{eq:new7}) holds, which indicates that the fourth-order lattice velocity tensor satisfies isotropy.

In addition, to satisfy the internal weighted orthogonality of $\mathbf H_3$, the following relations hold
\begin{subequations}\label{eq:new9}
\begin{equation}\label{eq:new9a}
(\mathbf{H}_3)_{\alpha\alpha\alpha} \mathbf W(\mathbf{H}_3)_{\alpha\beta\beta}=0, \quad \alpha \neq \beta,
\end{equation}
\begin{equation}\label{eq:new9b}
(\mathbf{H}_3)_{\alpha\alpha\beta} \mathbf W(\mathbf{H}_3)_{\beta\gamma\gamma}=0, \quad \text{$\alpha$, $\beta$, $\gamma$ are not equal to each other}.
\end{equation}
\end{subequations}
Combining Eq. (\ref{eq:new9}) with Eq. (\ref{eq:new6}) yields
\begin{subequations}\label{eq:new10}
\begin{equation}\label{eq:new10a}
\Delta_{\alpha\alpha\alpha\alpha\beta\beta}^{(6)}=\Delta_{\alpha\alpha\alpha\alpha}^{(4)}\Delta_{\beta\beta}^{(2)}, \quad \alpha \neq \beta,
\end{equation}
\begin{equation}\label{eq:new10b}
\Delta_{\alpha\alpha\beta\beta\gamma\gamma}^{(6)}=\Delta_{\alpha\alpha}^{(2)}\Delta_{\beta\beta}^{(2)}\Delta_{\gamma\gamma}^{(2)}, \quad \text{$\alpha$, $\beta$, $\gamma$ are not equal to each other}.
\end{equation}
\end{subequations}

Based on the above discussion on the weighted orthogonality of $\mathbf H_0, \mathbf H_1, \mathbf H_2,$ and $\mathbf H_3$, the weight coefficients are determined by the following even-order moments.
\begin{subequations}\label{eq:Cont1-5}
\begin{equation}
\Delta_{\alpha\alpha}^{(2)}=c_{s\alpha}^2,
\end{equation}
\begin{equation}
\begin{aligned}
\Delta_{\alpha\alpha\beta\beta}^{(4)}&=\Delta_{\alpha\alpha}^{(2)}\Delta_{\beta\beta}^{(2)}, \quad \alpha \neq \beta,
\end{aligned}
\end{equation}
\begin{equation}
\begin{aligned}
\Delta_{\alpha\alpha\alpha\alpha}^{(4)}&=3\Delta_{\alpha\alpha}^{(2)}\Delta_{\alpha\alpha}^{(2)},
\end{aligned}
\end{equation}
\begin{equation}
\begin{aligned}
\Delta_{\alpha\alpha\alpha\alpha\beta\beta}^{(6)}&=\Delta_{\alpha\alpha\alpha\alpha}^{(4)}\Delta_{\beta\beta}^{(2)}, \quad \alpha \neq \beta,
\end{aligned}
\end{equation}
\begin{equation}
\begin{aligned}
\Delta_{\alpha\alpha\beta\beta\gamma\gamma}^{(6)}&=\Delta_{\alpha\alpha}^{(2)}\Delta_{\beta\beta}^{(2)}\Delta_{\gamma\gamma}^{(2)}, \quad \text{$\alpha$, $\beta$, $\gamma$ are not equal to each other}.
\end{aligned}
\end{equation}
\end{subequations}

\emph{Remark 2.} (1) The lattice speed $c_{s\alpha}$ is taken as a parameter along $\alpha$-axis, as those in Refs. \cite{Ginzburg2005, dHumieres1992, Ladd1994, Ginzburg2003, Ginzburg2007}. Although $c_{s\alpha}$ is direction dependent, it is usually taken to be a direction-independent form with $c_{s\alpha}=c_s$ for simplicity. This is why we use $c_s$ in the DTE method.

(2) Eq. (\ref{eq:Cont1-5}a) and Eq. (\ref{eq:Cont1-5}b) imply that $\mathbf{H}_0$,  $\mathbf{H}_1$ and $\mathbf{H}_2$ are weighted orthogonal to each other and satisfying them gives a quadratic equilibrium state, and the related work can be found in Ref. \cite{ChaiShi2023}. Eq. (\ref{eq:Cont1-5}c) with Eq. (\ref{eq:Cont1-5}b) means that the weighted orthogonality between $\mathbf{H}_1$ and $\mathbf{H}_3$, which ensures that the fourth-order lattice velocity tensor is isotropic. The conditions of internal orthogonality of $\mathbf{H}_3$ are Eqs. (\ref{eq:Cont1-5}d) and (\ref{eq:Cont1-5}e), where Eq. (\ref{eq:Cont1-5}e) is only used in cases where the dimension $d$ is larger than 2.

Now, we focus on the $3$-layer velocity lattice model rD$d$Q($q+2\bar{q}$), where $\bar q=q-1$,
then for the velocity set \{$\mathbf 0, \mathbf{c}_j$, $2\mathbf{c}_{j}$, $3\mathbf{c}_j$, $1 \leq j\leq \bar{q}$\}, we have
\begin{equation}\label{eq:new1}
\Delta_{i_1 i_2 \cdots i_k}^{(k)}={c}_{i_1}{c}_{i_2}\cdots {c}_{i_k} \sum_{j =1}^{\bar q}(\omega_j+2^k\omega_{j+\bar q}+3^k \omega_{j+2\bar q}) \mathbf{e}_{ji_1}\mathbf{e}_{ji_2}\cdots \mathbf{e}_{ji_k}, k \geq 1,
\end{equation}
where $\mathbf{c}_{j\alpha}=c_{\alpha}\mathbf{e}_{j\alpha}$, $\alpha=1, \cdots, d$.  \{$\mathbf{c}_j, ~0 \leq j\leq q$\} is a standard velocity set. From Eq. (\ref{eq:new1}) one can obtain
\begin{subequations}\label{eq:new2a}
\begin{equation}\label{eq:new2}
\Delta_{\alpha\beta}^{(2)}={c}_{\alpha}{c}_{\beta}\sum_{j =1}^{\bar q}(\omega_j+2^2\omega_{j+\bar q}+3^2 \omega_{j+2\bar q}) \mathbf{e}_{j\alpha}\mathbf{e}_{j\beta},
\end{equation}
\begin{equation}\label{eq:new3}
\begin{aligned}
\Delta_{\alpha\alpha\beta\beta}^{(4)}&={c}_{\alpha}^2{c}_{\beta}^2\sum_{j =1}^{\bar q}(\omega_j+2^4\omega_{j+\bar q}+3^4 \omega_{j+2\bar q}) \mathbf{e}_{j\alpha}^2\mathbf{e}_{j\beta}^2,\\
& \xlongequal {\alpha=\beta}{c}_{\alpha}^4\sum_{j =1}^{\bar q}(\omega_j+2^4\omega_{j+\bar q}+3^4 \omega_{j+2\bar q}) \mathbf{e}_{j\alpha}^2,
\end{aligned}
\end{equation}
\begin{equation}\label{eq:new4}
\begin{aligned}
\Delta_{\alpha\alpha\beta\beta\gamma\gamma}^{(6)}&={c}_{\alpha}^2{c}_{\beta}^2{c}_{\gamma}^2\sum_{j =1}^{\bar q}(\omega_j+2^6\omega_{j+\bar q}+3^6 \omega_{j+2\bar q}) \mathbf{e}_{j\alpha}^2\mathbf{e}_{j\beta}^2\mathbf{e}_{j\gamma}^2,\\
&\xlongequal {\alpha=\gamma}{c}_{\alpha}^4 {c}_{\beta}^2\sum_{j =1}^{\bar q}(\omega_j+2^6\omega_{j+\bar q}+3^6 \omega_{j+2\bar q}) \mathbf{e}_{j\alpha}^2\mathbf{e}_{j\beta}^2,\\
&\xlongequal {\alpha=\beta=\gamma}{c}_{\alpha}^6 \sum_{j =1}^{\bar q}(\omega_j+2^6\omega_{j+\bar q}+3^6 \omega_{j+2\bar q}) \mathbf{e}_{j\alpha}^2,
\end{aligned}
\end{equation}
\end{subequations}
where $\mathbf{e}_{j\alpha}^m=\mathbf{e}_{j\alpha}$ ($m$ is odd) and $\mathbf{e}_{j\alpha}^m=\mathbf{e}_{j\alpha}^2$ ($m$ is even) are used for $1 \leq j \leq \bar{q} $.

Based on Eqs. (\ref{eq:Cont1-5}) - (\ref{eq:new2a}), the relationships satisfied by the weight coefficients can be expressed as
\begin{subequations}\label{eq:new11}
\begin{equation}\label{eq:new11a}
\Delta^{(0)}=\sum_j \omega_j=1,
\end{equation}
\begin{equation}\label{eq:new11b}
\Delta_{\alpha\alpha}^{(2)}=c_{s\alpha}^2 \Rightarrow \sum_{j =1}^{\bar q}(\omega_j+2^2\omega_{j+\bar q}+3^2 \omega_{j+2\bar q}) \mathbf{e}_{j\alpha}^2=d_{0\alpha},
\end{equation}
\begin{equation}\label{eq:new11c}
\Delta_{\alpha\alpha\beta\beta}^{(4)}=\Delta_{\alpha\alpha}^{(2)}\Delta_{\beta\beta}^{(2)} \Rightarrow \sum_{j =1}^{\bar q}(\omega_j+2^4\omega_{j+\bar q}+3^4 \omega_{j+2\bar q}) \mathbf{e}_{j\alpha}^2 \mathbf{e}_{j\beta}^2=d_{0\alpha}d_{0\beta}, \alpha \neq \beta,
\end{equation}
\begin{equation}\label{eq:new11d}
\Delta_{\alpha\alpha\alpha\alpha}^{(4)}=3\Delta_{\alpha\alpha}^{(2)} \Rightarrow \sum_{j =1}^{\bar q}(\omega_j+2^4\omega_{j+\bar q}+3^4 \omega_{j+2\bar q}) \mathbf{e}_{j\alpha}^2=3d_{0\alpha}^2,
\end{equation}
\begin{equation}\label{eq:new11e}
\begin{aligned}
\Delta_{\alpha\alpha\alpha\alpha\beta\beta}^{(6)}=\Delta_{\alpha\alpha\alpha\alpha}^{(4)}\Delta_{\beta\beta}^{(2)} & \Rightarrow \sum_{j =1}^{\bar q}(\omega_j+2^6\omega_{j+\bar q}+3^6 \omega_{j+2\bar q}) \mathbf{e}_{j\alpha}^2 \mathbf{e}_{j\beta}^2=3d_{0\alpha}^2 d_{0\beta}, \alpha \neq \beta,\\
\text{Exchange of $\alpha$, $\beta$}& \Rightarrow \sum_{j =1}^{\bar q}(\omega_j+2^6\omega_{j+\bar q}+3^6 \omega_{j+2\bar q}) \mathbf{e}_{j\alpha}^2 \mathbf{e}_{j\beta}^2=3d_{0\beta}^2 d_{0\alpha}\Rightarrow d_{0\alpha}=d_{0\beta},
\end{aligned}
\end{equation}
\begin{equation}\label{eq:new11f}
\begin{aligned}
\Delta_{\alpha\alpha\beta\beta\gamma\gamma}^{(6)}=\Delta_{\alpha\alpha}^{(2)}\Delta_{\beta\beta}^{(2)}\Delta_{\gamma\gamma}^{(2)} & \Rightarrow \sum_{j =1}^{\bar q}(\omega_j+2^6\omega_{j+\bar q}+3^6 \omega_{j+2\bar q}) \mathbf{e}_{j\alpha}^2 \mathbf{e}_{j\beta}^2 \mathbf{e}_{j\beta}^2=d_{0\alpha}d_{0\beta}d_{0\gamma}, \\
&\text{$\alpha$, $\beta$, $\gamma$ are not equal to each other},
\end{aligned}
\end{equation}
\end{subequations}
where $d_{0\alpha}=c_{s\alpha}^2/c_{\alpha}^2,~\alpha=1, \cdots, d$.

Eq. (\ref{eq:new11d})-Eq. (\ref{eq:new11b}) gives
\begin{equation}\label{eq:new12}
\sum_{j =1}^{\bar q}\left[(2^4-2^2)\omega_{j+\bar q}+(3^4-3^2) \omega_{j+2\bar q}\right] \mathbf{e}_{j\alpha}^2 =d_{0\alpha}(3d_{0\alpha}-1).
\end{equation}
Similarly, Eq. (\ref{eq:new11e})-Eq. (\ref{eq:new11c}) yields
\begin{equation}\label{eq:new13}
\sum_{j =1}^{\bar q}\left[(2^6-2^4)\omega_{j+\bar q}+(3^6-3^4) \omega_{j+2\bar q}\right] \mathbf{e}_{j\alpha}^2 \mathbf{e}_{j\beta}^2 =d_{0\alpha}d_{0\beta}(3d_{0\alpha}-1)=d_{0\alpha}d_{0\beta}(3d_{0\beta}-1).
\end{equation}

\emph{Remark 3.} When the lattice model is specified, the weighting coefficients can be solved from the above equations. If Eq. (\ref{eq:new11e}) or (\ref{eq:new13}) holds, then $d_{0\alpha}=d_{0\beta}$. If it also satisfies $c_{s\alpha}=c_s$, $\forall \alpha$, it follows that there is only a square lattice at this point. If only Eqs. (\ref{eq:new11b})-(\ref{eq:new11d})  are considered, then the third order EDF can be constructed and the isotropy condition can hold. Consider only that Eqs. (\ref{eq:new11a})-(\ref{eq:new11d}) hold, then \{$\mathbf{H}_0$, $\mathbf{H}_1$, $\mathbf{H}_2$, $\mathbf{H}_3$\} are block-weighted orthogonal, and the Hermite matrices, $\mathbf{H}_1$ and $\mathbf{H}_2$, are internally orthogonal except for $\mathbf{H}_3$. Therefore, we need to correct $\mathbf{H}_3$ to make it internally orthogonal.

To simplify the following analysis, we introduce $c_\alpha=\Delta x_{\alpha}/\Delta t$ ($\alpha=1, 2, \ldots, d$) in $d$-dimensional space with $\Delta x_{\alpha}$ being the spacing step in $\alpha$ axis. In this case, the discrete velocities and weight coefficients in the common rD$d$Q$b$ lattice models can be given. In the following, we only discuss the rD2Q25 lattice model. The discussion of other lattice models, rD3Q53 and rD2Q21 lattice models, can be found in Appendix.

rD2Q25 lattice:

\begin{equation}\label{eq:3-1}
\scriptsize{
\begin{split}
& \{ \mathbf{c}_j,0\leq j \leq 24\}=\\
&\left(
\begin{array}{ccccccccccccccccccccccccc}
    0 & c_1 &   0 & -c_1  &    0 & c_1 & -c_1 & -c_1 &  c_1&2 c_1 &   0 & -2c_1  &    0 & 2c_1 & -2c_1 & -2c_1 &  2c_1& 3c_1 &   0 & -3c_1  &    0 & 3c_1 & -3c_1 & -3c_1 &  3c_1 \\
    0 &   0 & c_2 &    0  & -c_2 & c_2 &  c_2 & -c_2 & -c_2&   0 & 2c_2 &    0  & -2c_2 & 2c_2 &  2c_2 & -2c_2 & -2c_2 &   0 &3 c_2 &    0  & -3c_2 & 3c_2 &  3c_2 & -3c_2 & -3c_2\\
\end{array}
\right),\\
\end{split}}
\end{equation}

\begin{equation}\label{eq:3-1-2}
\begin{split}
&\omega_j\geq 0, \omega_1=\omega_3,\omega_2=\omega_4,\omega_5=\omega_6=\omega_7=\omega_8,\omega_9=\omega_{11}, \omega_{10}=\omega_{12}, \\ &\omega_{13}=\omega_{14}=\omega_{15}=\omega_{16},\omega_{17}=\omega_{19},\omega_{18}=\omega_{20},\omega_{21}=\omega_{22}=\omega_{23}=\omega_{24},\omega_0=1-\sum_{j>0} \omega_j.
\end{split}
\end{equation}


From Eqs. (\ref{eq:new11}b)-(\ref{eq:new11}e), or Eqs. (\ref{eq:new11}b), (\ref{eq:new11}c), (\ref{eq:new12}), and (\ref{eq:new13}), one can obtain
\begin{subequations}
\begin{equation}\label{eq:3-3-3-1}
2(\omega_1+2\omega_5+4\omega_9+8\omega_{13}+9\omega_{17}+18\omega_{21})=d_{01},
\end{equation}
\begin{equation}\label{eq:3-3-3-2}
2(\omega_2+2\omega_5+4\omega_{10}+8\omega_{13}+9\omega_{18}+18\omega_{21})=d_{02},
\end{equation}
\begin{equation}\label{eq:3-3-3-3}
4\omega_5+64\omega_{13}+324\omega_{21}=d_{01}d_{02},
\end{equation}
\begin{equation}\label{eq:3-3-3-4}
2\omega_1+4\omega_5+32\omega_9+64\omega_{13}+162\omega_{17}+324\omega_{21}=3d_{01}^2,
\end{equation}
\begin{equation}\label{eq:3-3-3-5}
2\omega_2+4\omega_5+32\omega_{10}+64\omega_{13}+162\omega_{18}+324\omega_{21}=3d_{02}^2,
\end{equation}
\begin{equation}\label{eq:3-3-3-6}
4\omega_5+256\omega_{13}+2916\omega_{21}=3d_{01}^2 d_{02}=3d_{02}^2 d_{01},
\end{equation}
\end{subequations}
or equivalently
\begin{subequations}\label{eq:WCondQ25}
\begin{equation}
2(\omega_1+2\omega_5+4\omega_9+8\omega_{13}+9\omega_{17}+18\omega_{21})=d_{01},
\end{equation}
\begin{equation}
2(\omega_2+2\omega_5+4\omega_{10}+8\omega_{13}+9\omega_{18}+18\omega_{21})=d_{02},
\end{equation}
\begin{equation}
4\omega_5+64\omega_{13}+324\omega_{21}=d_{01}d_{02},
\end{equation}
\begin{equation}
24\omega_9+48\omega_{13}+144\omega_{17}+288\omega_{21}=d_{01}(3d_{01}-1),
\end{equation}
\begin{equation}
24\omega_{10}+48\omega_{13}+144\omega_{18}+288\omega_{21}=d_{02}(3d_{02}-1),
\end{equation}
\begin{equation}
4\omega_5+256\omega_{13}+2916\omega_{21}=3d_{01}^2 d_{02}=3d_{02}^2 d_{01}.
\end{equation}
\end{subequations}
Taking $\omega_{13}, \omega_{17}, \omega_{18}$, and $\omega_{21}$ as free weight coefficients, we obtain from Eqs. (\ref{eq:WCondQ25}a)-(\ref{eq:WCondQ25}e)
\begin{subequations}\label{eq:WeightQ25}
\begin{equation}
\omega_{5}=\frac{1}{4}\left( d_{01}d_{02}-64\omega_{13}-324\omega_{21}\right),
\end{equation}
\begin{equation}
\omega_{9}=\frac{1}{24}\left[ d_{01}(3d_{01}-1)-48\omega_{13}-144\omega_{17}-288\omega_{21}\right],
\end{equation}
\begin{equation}
\omega_{10}=\frac{1}{24}\left[ d_{02}(3d_{02}-1)-48\omega_{13}-144\omega_{18}-288\omega_{21}\right],
\end{equation}
\begin{equation}
\omega_{1}=\frac{1}{2}d_{01}-(2\omega_5+4\omega_9+8\omega_{13}+9\omega_{17}+18\omega_{21}),
\end{equation}
\begin{equation}
\omega_{2}=\frac{1}{2}d_{02}-(2\omega_5+4\omega_{10}+8\omega_{13}+9\omega_{18}+18\omega_{21}),
\end{equation}
\end{subequations}
which is equivalent to
\begin{subequations}\label{eq:WeightQ25-1}
\begin{equation}
\omega_{5}=\frac{1}{4}\left( d_{01}d_{02}-64\omega_{13}-324\omega_{21}\right),
\end{equation}
\begin{equation}
\omega_{17}=\frac{1}{144}\left[ d_{01}(3d_{01}-1)-24\omega_{9}-48\omega_{13}-288\omega_{21}\right],
\end{equation}
\begin{equation}
\omega_{18}=\frac{1}{144}\left[ d_{02}(3d_{02}-1)-24\omega_{10}-48\omega_{13}-288\omega_{21}\right],
\end{equation}
\begin{equation}
\omega_{1}=\frac{1}{2}d_{01}-(2\omega_5+4\omega_9+8\omega_{13}+9\omega_{17}+18\omega_{21}),
\end{equation}
\begin{equation}
\omega_{2}=\frac{1}{2}d_{02}-(2\omega_5+4\omega_{10}+8\omega_{13}+9\omega_{18}+18\omega_{21}),
\end{equation}
\end{subequations}
where $\omega_{9}, \omega_{10}, \omega_{13}$, and $\omega_{21}$ are free weight coefficients.

\emph{Remark 4.} As mentioned in \emph{Remark 3}, Eq. (\ref{eq:WeightQ25}) or Eq. (\ref{eq:WeightQ25-1}) gives the weights in rD2Q25 lattice model using the cubic EDF, and $\mathbf H_3$ needs to be corrected. If Eq. (\ref{eq:WCondQ25}f) is added to Eq. (\ref{eq:WeightQ25}) or Eq. (\ref{eq:WeightQ25-1}), we have $d_{01}=d_{02}$ ($=d_0$), and $\omega_{13}=(d_0^2(3d_0-1)-2592\omega_{21})/192$, which implies that there are no rectangular lattice when $c_{s1}=c_{s2}$, and the isotropy condition can hold for multispeed lattice (D2Q25) model.

For the two-dimensional case, $\mathbf H_3 = \{H_{xyy}, H_{xxy}, H_{xxx}, H_{yyy}\} $. If Eq. (\ref{eq:new13}) or Eq. (\ref{eq:WCondQ25}f) is not satisfied, the 'vectors' in $\mathbf H_3 $ are not weighted orthogonal and require correction. Furthermore, due to lattice symmetry, only two pairs, $\{H_{xyy}, H_{xxx}\} $ and $\{H_{xxy}, H_{yyy}\}$, are not weighted orthogonal. This means it is sufficient to correct $H_{xxx}$ and $H_{yyy}$, respectively. Thus, we have
\begin{subequations}\label{eq:CorrectQ25}
\begin{equation}
H_{xxx}:=H_{xxx}-\frac{H_{xxx}\mathbf W H_{xyy}^{T}}{H_{xyy}\mathbf W H_{xyy}^{T}}H_{xyy},
\end{equation}
\begin{equation}
H_{yyy}:=H_{yyy}-\frac{H_{yyy}\mathbf W H_{xxy}^{T}}{H_{xxy}\mathbf W H_{xxy}^{T}}H_{xxy},
\end{equation}
\end{subequations}
where $H_{\alpha\beta\gamma}$ on the right side of Eq. (\ref{eq:CorrectQ25}) defined as
\begin{equation}\label{eq:CorrectQ25-1}
H_{xyy}=c_{jx}(c_{jy}^2-c_{s2}^2),~H_{xxy}=c_{jy}(c_{jx}^2-c_{s1}^2),~H_{xxx}=c_{jx}(c_{jx}^2-3c_{s1}^2),~H_{yyy}=c_{jy}(c_{jy}^2-3c_{s2}^2).
\end{equation}

Based on Eqs. (\ref{eq:WeightQ25}), (\ref{eq:CorrectQ25}), and (\ref{eq:new11}a), or Eqs. (\ref{eq:WeightQ25-1}), (\ref{eq:CorrectQ25}), and (\ref{eq:new11}a), several lattice models that can be obtained as the subsets of the rD2Q25 lattice by setting some weight coefficients to zero and removing the corresponding velocities in the rD2Q25 lattice, where the weights need to be relabeled.

(i) rD2Q21 lattice: $\omega_{21}=0$; $\omega_{13}=0$; $\omega_{17}=\omega_{18}=0$; $\omega_9=\omega_{10}=0$; etc.

(ii) rD2Q17 lattice: $\omega_{21}=\omega_{9}=\omega_{10}=0$; $\omega_{13}=\omega_{9}=\omega_{10}=0$; $\omega_{13}=\omega_{1}=\omega_{2}=0$; $\omega_9=\omega_{10}=\omega_{1}=\omega_{2}=0$; $\omega_{21}=\omega_{17}=\omega_{18}=0$; $\omega_{13}=\omega_{17}=\omega_{18}=0$; $\omega_{21}=\omega_{13}=0$;  etc. Details of the special cases of the first five rD2Q17 lattice models can be found in Appendix~\ref{app:sec1}.

(iii) rD2Q13 lattice: $\omega_{21}=\omega_{13}=\omega_{17}=\omega_{18}=0$; $\omega_{21}=\omega_{13}=\omega_{9}=\omega_{10}=0$; etc.

Here, the first rD2Q13 lattice above is given as follows.
\begin{equation}\label{eq:Q13lattice}
\begin{split}
&\{\mathbf{c}_j,0\leq j \leq 12\}=
\left(
\begin{array}{ccccccccccccc}
    0 & c_1 &   0 & -c_1  &    0 & c_1 & -c_1 & -c_1 &  c_1 & 2c_1 &   0      & -2c_1 &  0   \\
    0 &   0 & c_2 &    0  & -c_2 & c_2 &  c_2 & -c_2 & -c_2 &   0     &  2c_2 &   0     & -2c_2 \\
\end{array}
\right),\\
& d_{01}=c_{s1}^2/{c_1^2},\quad d_{02}=c_{s2}^2/{c_2^2},\\
&\omega_{9}=d_{01}(3d_{01}-1)/24,\quad \omega_{10}=d_{02}(3d_{02}-1)/24,\quad \omega_0=(1-d_{01})(1-d_{02})+6(\omega_{9}+\omega_{10}),\\ &\omega_1=d_{01}(1-d_{02})/2-4\omega_9,  \quad \omega_2=d_{02}(1-d_{01})/2-4\omega_{10},\quad \omega_5=d_{01}d_{02}/4,\\
\end{split}
\end{equation}
where $1/3<d_{01},d_{02}<1, 2/3<d_{01}+d_{02}<4/3$. Then the weight matrix $\mathbf W$ can be written as
\begin{equation}
\mathbf W=\text{diag}\{\omega_0,\omega_1,\omega_2,\omega_1,\omega_2,\omega_5,\omega_5,\omega_5,\omega_5,\omega_9,\omega_{10},\omega_9,\omega_{10}\}.
\end{equation}
and the transformation matrix is
\begin{equation}\label{eq:TransM}
\mathbf M=\mathbf D
\left(
\begin{array}{ccccccccccccc}
   1 & 1 &   1 & 1 & 1 &   1&1 & 1 &   1 & 1 & 1 &   1& 1   \\
    0 & 1 &   0 & -1  &    0 & 1 & -1 & -1 &  1 & 2 &   0   & -2 &  0   \\
    0 &  0 & 1 &    0  & -1 & 1 &  1 & -1 & -1 &   0     &  2 &   0    & -2 \\
    0 & 1 &   0 &  1  &    0 & 1 & 1 & 1 &  1 & 4 &   0   & 4 &  0   \\
    0 & 0 &   0 &  0  &    0 & 1 & -1 & 1 &  -1 & 0 &   0   & 0 &  0   \\
    0 &  0 & 1 &    0  & 1  & 1 &  1 & 1 & 1 &   0     &  4 &   0    & 4 \\
    0 &  0 & 0 &    0  & 0 & 1 &  -1 & -1 & 1 &   0     &  0 &   0    & 0 \\
    0 & 0 &   0 &  0  &    0 & 1 & 1 & -1 &  -1 & 0 &   0   & 0 &  0   \\
    0 & 1 &   0 & -1  &    0 & 1 & -1 & -1 &  1 & 8 &   0   & -8 &  0   \\
    0 &  0 & 1 &    0  & -1 & 1 &  1 & -1 & -1 &   0     &  8 &   0    & -8\\
    0 & 1 &   0 & 1  &    0 & 1 & 1 & 1 &  1 & 16 &   0   & 16 &  0   \\
    0 & 0 &   0 & 0  &    0 & 1 & 1 & 1 &  1 & 0 &   0   & 0 &  0   \\
    0 &  0 & 1 &    0  & 1 & 1 &  1 & 1 & 1 &   0     &  16 &   0    & 16 \\
\end{array}
\right),\\
\end{equation}
where
\begin{equation}\label{eq:TransM-D}
\mathbf D=\text{diag}\{1,c_1,c_2,c_1^2,c_1c_2,c_2^2,c_1c_2^2,c_1^2c_2,c_1^3,c_2^3,c_1^4,c_1^2c_2^2,c_2^4\}.
\end{equation}

\emph{Remark 5.} (1) When Eq. (\ref{eq:WCondQ25}f) is considered, the rD2Q13 lattice models cannot be obtained. This is due to the fact that when the rD2Q13 lattice model is considered, the weight coefficients $\omega_{21}=\omega_{13}$, which leads to $d_{01}=d_{02}=d_0=1/3$ from Eqs. (\ref{eq:WCondQ25}c) and (\ref{eq:WCondQ25}f). Then it follows from Eqs. (\ref{eq:WCondQ25}d) and (\ref{eq:WCondQ25}e) that $\omega_{17}=\omega_{18}=0$ if and only if $\omega_{9}=\omega_{10}=0$, which implies that the rD2Q13 lattice models do not exist.

(2) One can also obtain the asymmetric sub-lattices of the rD2Q25 lattice, just as Hegeler et al. did in Ref. \cite{Hegeler2013}. For example, rD2Q11 lattice with $\omega_{21}=\omega_{13}=\omega_{17}=\omega_{18}=0$ and $\omega_9=0$ ($d_{01}=1/3$) or $\omega_{10}=0 $ ($d_{02}=1/3)$, and rD2Q15 with $\omega_{21}=\omega_{17}=\omega_{18}=0$ and $\omega_9=0$ or $\omega_{10}=0 $.

\section{Conclusions}

In this work, we present a unified multi-relaxation-time lattice Boltzmann (MRT-LB) framework for the Navier-Stokes equations (NSEs) and nonlinear convection-diffusion equations (NCDE) based on Hermite matrices and multispeed rectangular lattice (rDdQb) models. Key contributions include:

\textbf{Unified MRT-LB Framework}: We establish a generalized MRT-LB framework capable of simulating both incompressible and compressible flows in single-phase and multiphase systems, as well as nonlinear convection-diffusion phenomena. The framework is derived via direct discrete modeling (DDM) from the MRT discrete-velocity Boltzmann equation (MRT-DVBE) and MRT lattice Boltzmann method (MRT-LBM).

\textbf{Moment Equations and Target Equations}: By using direct Taylor expansion analysis, we derive macroscopic moment equations from the MRT-LB model and MRT-DVBE. The recovery of the target NSEs and NCDE relies on the proper selection of fundamental moments and the introduction of auxiliary moments. These auxiliary moments play a crucial role in eliminating spurious terms and recovering the correct macroscopic physics.

\textbf{Hermite Matrices and Multispeed Lattices}: Using the weighted orthogonality of Hermite matrices, we construct several multispeed rectangular lattice models, including rD2Q25, rD2Q21, rD2Q17, rD2Q13, rD3Q53, rD3Q45, and rD3Q33. A generalized third-order equilibrium distribution function is derived, and corrections are applied to specific elements of the third-order Hermite matrix to ensure orthogonality on rectangular lattices.

\textbf{Flexibility and Extensibility}: The proposed framework generalizes and extends previous MRT-LB models, offering a systematic approach for constructing LB models on standard and non-standard lattices. The methodology is not limited to isothermal or weakly compressible flows, making it applicable to a broader range of fluid dynamics and transport problems.

While this study provides a comprehensive and unified MRT-LB framework, several directions remain for future exploration:

\textbf{Extension to higher-order MRT-LB model with higher-order Hermite expansions} for improving the accuracy and stability of the LBM.

\textbf{Further theoretical and numerical analysis} of the proposed multispeed lattice models, such as the stability and boundary treatment, as well as the effect of free parameters: free relaxation factors, moments, etc.

\textbf{Application to multiphysics problems}, such as thermal flows, reactive flows, and multiphase systems with complex interfacial dynamics.

\textbf{Development of other efficient MRT-LBE methods}, including MRT-FDLBM, MRT-FVLBM etc.


\section*{Acknowledgements}
This work was financially supported by the National Natural Science Foundation of China (Grants No. 12072127 and No. 12202130).

\appendix
\section{\label{app:sec3}The derivation of the moment equation for nonlinear convection-diffusion equation using DTE method}

First we derive the moment equation for NCDE from MRT-LB method (\ref{eq:2-1}). Here we give the following basic moments on $\mathbf{\Lambda}$, $f_j$, $f_j^{eq}$, $G_j$, and $F_j$,
\begin{subequations}\label{eq:M-NCDE}
\begin{equation}\label{eq:M-NCDE-1a}
M_0=\sum_j f_j^{eq}=\sum_j f_j+\frac{\Delta t \lambda}{2} S_G(\mathbf{x}, t),\ \  \mathbf{M}_1=\sum_j \mathbf{c}_j f_j^{eq}, \mathbf{M}_2=\sum_j \mathbf{c}_j \mathbf{c}_j f_j^{eq},
\end{equation}
\begin{equation}
M_{0F}=\sum_j F_j,\ \ \mathbf{M}_{1F}=\sum_j \mathbf{c}_j F_j,
\end{equation}
\begin{equation}
M_{0G}=\sum_j G_j,\ \ \mathbf{M}_{1G}=\sum_j \mathbf{c}_j G_j,
\end{equation}
\begin{equation}
\sum_j\mathbf{e}_j\mathbf{\Lambda}_{jk} = s_0\mathbf{e}_k, \ \  \sum_j \mathbf{c}_j\mathbf{\Lambda}_{jk} = \mathbf{S}_{10} \mathbf{e}_k + \mathbf{S}_1 \mathbf{c}_k,
\end{equation}
\end{subequations}
where $\mathbf{M}_k$, $\mathbf{M}_{kF}$ and $\mathbf{M}_{kG}$ ($k\geq 0$) are the $k$-th moments of $f_j^{eq}, F_j$ and $G_j$, respectively.
 $\mathbf{S}_{10}$ is a $d\times 1$ matrix, $\mathbf{S}_1$ is an invertible $d\times d$ relaxation sub-matrix corresponding to the diffusion tensor. Additionally, the first equation in Eq. (\ref{eq:M-NCDE}a) gives the following moment of nonequilibrium,
\begin{equation}\label{eq:M-NCDE-1}
M_0^{ne}=\sum_j f_j^{ne}=\sum_j (f_j-f_j^{eq})=-\frac{\Delta t \lambda}{2} S_{G}(\mathbf{x}, t).
\end{equation}

Summing Eq. (\ref{eq:4-3}a) and Eq. (\ref{eq:4-5}), and adopting Eqs. (\ref{eq:M-NCDE}) and (\ref{eq:M-NCDE-1}), one can obtain
\begin{subequations}\label{eq:ME-NCDE}
\begin{eqnarray}
\partial_t M_0+\nabla\cdot \mathbf{M}_1 &=& -\frac{s_0}{\Delta t} M_0^{ne}+ M_{0G}+M_{0F}+O(\Delta t) \nonumber\\ &=& \frac{1}{2}s_0 \lambda S_{G}+ M_{0G}+M_{0F}+O(\Delta t),
\end{eqnarray}
\begin{eqnarray}
\partial_t M_0+\nabla\cdot \mathbf{M}_1 +\partial_t(1-s_0/2)M_0^{ne}+\nabla\cdot \big[(\mathbf{I-S}_1 /2)\mathbf{M}_1^{ne}-\mathbf{S}_{10}M_0^{ne}/2\big]\nonumber\\
 + \frac{\Delta t}{2} \partial_t(M_{0G}+(1-\theta)M_{0F})+\frac{\Delta t}{2} \nabla\cdot (\mathbf{M}_{1G}+(1-\gamma)\mathbf{M}_{1F}) \nonumber\\
= \frac{1}{2}s_0 \lambda S_{G}+ M_{0G}+M_{0F}+O(\Delta t^2),
\end{eqnarray}
\end{subequations}
where
\begin{equation}\label{eq:M-NCDE-2}
\mathbf{M}_1^{ne}=\sum_j \mathbf{c}_j f_j^{ne}=\sum_j \mathbf{c}_j(f_j-f_j^{eq}).
\end{equation}

Multiplying $\mathbf{c}_j$ on both sides of Eq. (\ref{eq:4-3}a), and through a summation over $j$, and using Eqs. (\ref{eq:M-NCDE}) and (\ref{eq:M-NCDE-1}), we have
\begin{eqnarray}\label{eq:M-NCDE-3-0}
\partial_t \mathbf{M}_1+\nabla\cdot \mathbf{M}_2=  -\frac{1}{\Delta t} (\mathbf{S}_1\mathbf{M}_1^{ne}+\mathbf{S}_{10}M_0^{ne})+ \mathbf{M}_{1G}+\mathbf{\mathbf{M}}_{1F} +O(\Delta t)\nonumber\\
=-\frac{1}{\Delta t} \mathbf{S}_1\mathbf{M}_1^{ne} +\frac{\lambda}{2}\mathbf{S}_{10}S_{G}+ \mathbf{M}_{1G}+\mathbf{\mathbf{M}}_{1F} +O(\Delta t),
\end{eqnarray}
then
\begin{equation}\label{eq:M-NCDE-3}
\mathbf{M}_1^{ne}=-\Delta t \mathbf{S}_1^{-1}(\partial_t \mathbf{M}_1+\nabla\cdot \mathbf{M}_2-\mathbf{M}_{1G}-\mathbf{\mathbf{M}}_{1F}-\frac{\lambda}{2}\mathbf{S}_{10}S_{G})+O(\Delta t^2).
\end{equation}

Substituting Eqs. (\ref{eq:M-NCDE-1}) and (\ref{eq:M-NCDE-3}) into Eq. (\ref{eq:ME-NCDE}b), we can obtain
\begin{eqnarray}\label{eq:ME-NCDE-1}
\partial_t M_0+\nabla\cdot \mathbf{M}_1 +\frac{\Delta t}{2}\partial_t\big[(1-\theta)M_{0F}+M_{0G}-(1-s_0/2)\lambda S_G\big]\nonumber\\
=\Delta t\nabla\cdot \big[ \mathbf{(S_1^{-1}-I}/2)(\partial_t \mathbf{M}_1+\nabla\cdot \mathbf{M}_2)-\mathbf{S}_1^{-1}(\mathbf{M}_{1G}+\mathbf{M}_{1F}+\frac{\lambda}{2}\mathbf{S}_{10}S_{G})+\frac{\gamma}{2}\mathbf{M}_{1F}\big]\nonumber\\
+\frac{1}{2}s_0 \lambda S_{G}+ M_{0G}+M_{0F}+O(\Delta t^2).
\end{eqnarray}

Let
\begin{equation}\label{eq:M-NCDE-4}
M_{0G}=(1-\frac{1}{2}s_0\lambda)S_G, M_{0F}=S_F, S=S_G+S_F,
\end{equation}
we obtain $\frac{1}{2}s_0 \lambda S_{G}+ M_{0G}+M_{0F}=S$, and it follows from Eqs. (\ref{eq:ME-NCDE}a) and (\ref{eq:ME-NCDE-1}) that
\begin{subequations}\label{eq:ME-NCDE-2}
\begin{equation}
\partial_t M_0+\nabla\cdot \mathbf{M}_1 = S+O(\Delta t),
\end{equation}
\begin{eqnarray}
\partial_t M_0+\nabla\cdot \mathbf{M}_1 +\frac{\Delta t}{2}\partial_t\big[(1-\theta)S_F+(1-\lambda)S_G\big]\nonumber\\
=\Delta t\nabla\cdot \big[ \mathbf{(S_1^{-1}-I}/2)(\partial_t \mathbf{M}_1+\nabla\cdot \mathbf{M}_2)-\mathbf{S}_1^{-1}(\mathbf{M}_{1G}+\mathbf{M}_{1F}+\frac{\lambda}{2}\mathbf{S}_{10}S_{G})+\frac{\gamma}{2}\mathbf{M}_{1F}\big]+S+O(\Delta t^2).
\end{eqnarray}
\end{subequations}

Taking the parameters and moments $\mathbf{M}_{1G}, \mathbf{M}_{1F}$ in Eq. (\ref{eq:ME-NCDE-2}) properly, we can obtain the expected moment equation for NCDE.
Let
\begin{equation}\label{eq:M-NCDE-5}
(1-\theta)S_F+(1-\lambda)S_G=0,\mathbf{M}_{1F}=0,
\end{equation}
Eq.(\ref{eq:ME-NCDE-2}) becomes
\begin{subequations}\label{eq:ME-NCDE-3}
\begin{equation}
\partial_t M_0+\nabla\cdot \mathbf{M}_1 = S+O(\Delta t),
\end{equation}
\begin{eqnarray}
\partial_t M_0+\nabla\cdot \mathbf{M}_1 =\Delta t\nabla\cdot \big[ {( \mathbf S_1^{-1}-\mathbf I}/2)(\partial_t \mathbf{M}_1+\nabla\cdot \mathbf{M}_2)-\mathbf{S}_1^{-1}\mathbf{\bar{M}}_{1G}\big]+S+O(\Delta t^2),
\end{eqnarray}
\end{subequations}
where $\mathbf{\bar{M}}_{1G}=\mathbf{M}_{1G}+\frac{\lambda}{2}\mathbf{S}_{10}S_{G}$, and the relaxation sub-matrix $\mathbf{S}_1$ and auxiliary moment $\mathbf{M}_{1G}$ need to be determined.

In the moment equation (\ref{eq:ME-NCDE-3}), taking
\begin{eqnarray}\label{eq:M-NCDE-6}
M_0=\phi, \mathbf{M}_1=\mathbf{B}, \mathbf{M}_2=c_s^2 \mathbf{D}+\mathbf{M}_{20},\nonumber\\ \mathbf{\bar{M}}_{1G}=(\mathbf{I}-\mathbf{S}_1/2)(\partial_t \mathbf{M}_1+\nabla\cdot \mathbf{M}_{20})-\mathbf{S}_1(\mathbf{A}_1\partial_t \mathbf{\bar{B}}+\mathbf{A}_2\nabla\cdot \mathbf{\bar{D}}),
\end{eqnarray}
one can obtain the following NCDE with a general form
\begin{equation}\label{eq:GNCDE}
\partial_t \phi+\nabla\cdot \mathbf{B} =\nabla\cdot \big[\mathbf{K} \nabla\cdot \mathbf{D}+\mathbf{K}_1 \partial_t\mathbf{\bar{B}}+\mathbf{K}_2 \nabla\cdot \mathbf{\bar{D}}\big]+S+O(\Delta t^2)
\end{equation}
with
\begin{equation}\label{eq:GNCDE-0}
\mathbf{K}= \Delta t c_s^2 (\mathbf{S}_1-\mathbf{I}/2), \mathbf{K}_1= \Delta t \mathbf{S}_1\mathbf{A}_1, \mathbf{K}_2= \Delta t \mathbf{S}_1\mathbf{A}_2.
\end{equation}

It should be noted that Eq. (\ref{eq:GNCDE}) is a more general NCDE with cross a diffusion term and a mixed partial derivative (More diffusion or mixed partial derivative terms can also be given). Some of its special cases can be derived from Eq. (\ref{eq:GNCDE}).

(1) NCDE without cross diffusion term and mixed partial derivative

$\mathbf{K}_1=0, \mathbf{K}_2=0, \mathbf{\bar{M}}_{1G}=(\mathbf{I}-\mathbf{S}_1^{-1}/2)(\partial_t \mathbf{M}_1+\nabla\cdot \mathbf{M}_{20})$.

(2) Diffusion equation

$\mathbf{M}_1=0, \mathbf{M}_{20}=0, \mathbf{K}_1=0, \mathbf{K}_2=0, \mathbf{\bar{M}}_{1G}=0$.

(3) Diffusion equation based MRT-LB model for convection-diffusion equation (CDE)

If we take $\mathbf{M}_1=0, \mathbf{M}_2=c_s^2 \phi \mathbf{I}, \mathbf{M}_{20}=0, \mathbf{K}_1=0, \mathbf{K}_2=0, \lambda =1, \mathbf{\bar{M}}_{1G}=0$, and $S_G=-\mathbf{u}\cdot\nabla \phi, S_F=S$, then the following CDE can be recovered from Eq. (\ref{eq:GNCDE}).
\begin{equation}\label{eq:CDE-1}
\partial_t \phi+\mathbf{u}\cdot\nabla \phi =\nabla\cdot \big[\mathbf{K} \nabla \phi\big]+S,
\end{equation}
where $\mathbf{u}$ is a known function of $\mathbf{x}$ and $t$. For this case, we get from Eqs. (\ref{eq:M-NCDE-3}) and (\ref{eq:2-4-0}a) or (\ref{eq:M-NCDE}a) that
\begin{subequations}\label{eq:M-NCDE-3-00}
\begin{equation}
\nabla \phi=-\frac{\mathbf{S}_1}{\Delta t c_s^2} \sum_j f_j+O(\Delta t),
\end{equation}
\begin{equation}
\phi = \sum_j f_j+ \frac{\Delta t \lambda}{2}S_G=(\mathbf{I}+\frac{\mathbf{uS}_1}{2 c_s^2})\sum_j f_j.
\end{equation}
\end{subequations}

It should be noted that Eq. (\ref{eq:CDE-1}) can also be treated as a CDE with the following form
\begin{equation}\label{CDE-2}
\partial_t \phi+\nabla \cdot (\mathbf{u}\phi) =\nabla\cdot \big[\mathbf{K} \nabla \phi\big]+S+\phi \nabla\cdot \mathbf{u}.
\end{equation}

\section{\label{app:sec0}The moment equations of MRT-DVBE and MRT-LBE}
The MRT-LB method [Eq. (\ref{eq:2-1})] can be obtained from the following discrete-velocity Boltzmann equation (DVBE)
\begin{equation}\label{app:1-1}
D_j f_j(\mathbf{x}, t)=-\bar {\mathbf{\Lambda}}_{jk} f_k^{ne}(\mathbf{x}, t)+ \bar G_j(\mathbf{x},t)+F_j(\mathbf{x}, t),
\end{equation}
where $\bar {\mathbf{\Lambda}}_{jk}$ represents the collision matrix, $\bar G_j(\mathbf{x}, t)$ is used to remove some additional terms, and $F_j(\mathbf{x}, t)$ is the source term. Integrating Eq. (\ref{app:1-1}) along the characteristic line $\bar{\mathbf x}=\mathbf x+\mathbf c_j \bar t$ with $\bar t \in [0,\Delta t]$, and using the trapezoidal formula and Taylor expansion for the right-hand term (see Ref. \cite{ChaiShi2020}), one can obtain
\begin{equation}\label{app:1-2}
 \bar f_j(\mathbf{x}+\mathbf{c}_j \Delta t,t+\Delta t)=\bar f_j(\mathbf{x}, t)-\mathbf{\Lambda}_{jk} \bar f_k^{ne}(\mathbf{x}, t)+\Delta t \big[G_j(\mathbf{x},t)+F_j(\mathbf{x}, t)+\frac{\Delta t}{2}\bar{D}_j F_j(\mathbf{x}, t)\big],
\end{equation}
with the following relation
\begin{equation}
\mathbf{\Lambda}=(\mathbf I/2+\bar{\mathbf{\Lambda}}^{-1}/{\Delta t})^{-1},\quad G_j=(\delta_{jk}-\mathbf{\Lambda}_{jk}/2)\bar G_k,
\quad \bar f_j=f_j-\frac{\Delta t}{2}\left( -\bar {\mathbf{\Lambda}}_{jk}f_k^{ne}+\bar G_j \right).
\end{equation}
Note that for simplicity,  we use $f_j$ for $\bar f_j$ in Eq. (\ref{eq:2-1}).

Based on the CE analysis (or DTE method), we can obtain the moment equations of the MRT-DVBE and MRT-LBM for the NCDE respectively,
\begin{subequations}
\begin{equation}
\partial_t M_0+\nabla\cdot \mathbf{M}_1 =\nabla\cdot  \bar {\mathbf S}_1^{-1} \big[ \partial_t \mathbf{M}_1+\nabla\cdot \mathbf{M}_2-\mathbf{M}_{1\bar G}\big]+M_{0F},
\end{equation}
\begin{equation}
\partial_t M_0+\nabla\cdot \mathbf{M}_1 =\Delta t\nabla\cdot \big[ {( \mathbf S_1^{-1}-\mathbf I}/2)(\partial_t \mathbf{M}_1+\nabla\cdot \mathbf{M}_2)-\mathbf{S}_1^{-1}\mathbf{M}_{1G}\big]+M_{0F},
\end{equation}
\end{subequations}
where the relations $\bar{\mathbf S}_1^{-1}=(\mathbf S_1^{-1}-\mathbf I/2)\Delta t$, $\mathbf{M}_{1G}=(\mathbf I-\mathbf S_1/2)\mathbf{M}_{1\bar G}$ hold, and $\lambda=0$ in Eq. (\ref{eq:M-NCDE-1a}). It is easy to obtain that the above two moment equations are consistent.

Similarly, through the CE analysis (or DTE method) one can also obtain the moment equations of MRT-DVBE and MRT-LBM for the NSEs respectively.

Moment equations of MRT-DVBE
\begin{subequations}
\begin{equation}
\partial_t M_0+\nabla\cdot \mathbf{M}_1 =M_{0F},
\end{equation}
\begin{equation}
\partial_t M_1+\nabla\cdot \mathbf{M}_2 =\mathbf M_{1F}+\mathbf {\nabla}\cdot \bar  {\mathbf S}_2^{-1}\big [ \partial _t \mathbf M_2+\mathbf {\nabla}\cdot \mathbf M_3-\mathbf{M}_{2\bar G}\big],
\end{equation}
\end{subequations}

Moment equations of MRT-LBM
\begin{subequations}
\begin{equation}
\partial_t M_0+\nabla\cdot \mathbf{M}_1 =M_{0F},
\end{equation}
\begin{equation}
\partial_t M_0+\nabla\cdot \mathbf{M}_1 =\mathbf M_{1F}+\Delta t\nabla\cdot \big[ {( \mathbf S_2^{-1}-\mathbf I}/2)(\partial_t \mathbf{M}_2+\nabla\cdot \mathbf{M}_3)-\mathbf{S}_2^{-1}\mathbf{M}_{2G}\big],
\end{equation}
\end{subequations}
where $\bar{\mathbf S}_2^{-1}=(\mathbf S_2^{-1}-\mathbf I/2)\Delta t$, $\mathbf{M}_{2G}=(\mathbf I-\mathbf S_2/2)\mathbf{M}_{2\bar G}$. It can be obtained that the moment equations of MRT-DVBE and MRT-LBM are also consistent.
\section{\label{app:sec1}Lattice models of RMRT-LB method on rectangular lattice in 2D}

In the 2D case, we will give some special cases for the rD2Q17 lattice model based on the rD2Q25 lattice where $d_{01}=d_{02}=d_0$, $c_{s1}=c_{s2}=c_s$ and $c_1=c_2=c=1$ are satisfied.

rD2Q17 lattice:

(1) Case 1
\begin{equation}
\begin{split}
&\{\mathbf{c}_j,0\leq j \leq 16\}=
\left(
\begin{array}{ccccccccccccccccc}
    0 & c_1 &   0 & -c_1  &    0 & c_1 & -c_1 & -c_1 &  c_1 & 2c_1 & -2c_1 & -2c_1 &  2c_1& 3c_1 &   0 & -3c_1  &    0  \\
    0 &   0 & c_2 &    0  & -c_2 & c_2 &  c_2 & -c_2 & -c_2 & 2c_2 &  2c_2 & -2c_2 & -2c_2 &   0 &3 c_2 &    0  & -3c_2 \\
\end{array}
\right),\\
&\omega_0=(575 + 193 \sqrt{193})/8100, \omega_1=(3355-91\sqrt{193})/18000,  \omega_5=(655+17 \sqrt{193})/27000,\\
&\omega_9=(685-49 \sqrt{193})/54000, \omega_{13}=(1445-101 \sqrt{193})/162000;c_s^2=72/(125+5 \sqrt{193}).
\end{split}
\end{equation}
Then the weight matrix $\mathbf W$ can be written as
\begin{equation}
\mathbf W=\text{diag}\{\omega_0,\omega_1,\omega_1,\omega_1,\omega_1,\omega_5,\omega_5,\omega_5,\omega_5,\omega_9,\omega_9,\omega_9,\omega_9,\omega_{13},\omega_{13},\omega_{13},\omega_{13}\}.
\end{equation}

(2) Case 2
\begin{equation}
\begin{split}
&\{\mathbf{c}_j,0\leq j \leq 16\}=
\left(
\begin{array}{ccccccccccccccccc}
    0 & c_1 &   0 & -c_1  &    0 & c_1 & -c_1 & -c_1 &  c_1 &  3c_1 &   0 & -3c_1  &    0 & 3c_1 & -3c_1 & -3c_1 &  3c_1 \\
    0 &   0 & c_2 &    0  & -c_2 & c_2 &  c_2 & -c_2 & -c_2 &   0  &3 c_2 &    0  & -3c_2 & 3c_2 &  3c_2 & -3c_2 & -3c_2\\
\end{array}
\right),\\
&\omega_0=(190 - 8 \sqrt{10})/ 405, \omega_1=(12 \sqrt{10}-15)/200,  \omega_5=(150-39\sqrt{10})/800,\\
&\omega_9=(295-92\sqrt{10})/162000, \omega_{13}=(130-41\sqrt{10})/648000;c_s^2=3/(5+\sqrt{10}).
\end{split}
\end{equation}
The weight matrix $\mathbf W$ is given by
\begin{equation}
\mathbf W=\text{diag}\{\omega_0,\omega_1,\omega_1,\omega_1,\omega_1,\omega_5,\omega_5,\omega_5,\omega_5,\omega_9,\omega_9,\omega_9,\omega_9,\omega_{13},\omega_{13},\omega_{13},\omega_{13}\}.
\end{equation}

(3) Case 3
\begin{equation}
\begin{split}
&\{\mathbf{c}_j,0\leq j \leq 16\}=
\left(
\begin{array}{ccccccccccccccccc}
    0 & c_1 & -c_1 & -c_1 &  c_1 & 2c_1 & 0 & -2c_1 &  0  &  3c_1 &   0 & -3c_1  &    0 & 3c_1 & -3c_1 & -3c_1 &  3c_1 \\
    0 & c_2 &  c_2 & -c_2 & -c_2 & 0  &2 c_2 &    0  & -2c_2 & 0  &3 c_2 &    0  & -3c_2 & 3c_2 &  3c_2 & -3c_2 & -3c_2\\
\end{array}
\right),\\
&\omega_0=455/1152, \omega_1=243/2048,  \omega_5=81/2560,\omega_9=1/1440, \omega_{13}=5/18432;c_s^2=3/4.
\end{split}
\end{equation}
The weight matrix $\mathbf W$ is given by
\begin{equation}
\mathbf W=\text{diag}\{\omega_0,\omega_1,\omega_1,\omega_1,\omega_1,\omega_5,\omega_5,\omega_5,\omega_5,\omega_9,\omega_9,\omega_9,\omega_9,\omega_{13},\omega_{13},\omega_{13},\omega_{13}\}.
\end{equation}

(4) Case 4
\begin{equation}
\begin{split}
&\{\mathbf{c}_j,0\leq j \leq 16\}=
\left(
\begin{array}{ccccccccccccccccc}
    0 & c_1 & -c_1 & -c_1 &  c_1 & 2c_1 & -2c_1 & -2c_1 &  2c_1  &  3c_1 &   0 & -3c_1  &    0 & 3c_1 & -3c_1 & -3c_1 &  3c_1 \\
    0 & c_2 &  c_2 & -c_2 & -c_2 & 2c_2  &2c_2  &   -2c_2  & -2c_2 & 0  &3 c_2 &    0  & -3c_2 & 3c_2 &  3c_2 & -3c_2 & -3c_2\\
\end{array}
\right),\\
&\omega_0=35/288, \omega_1=45/256,  \omega_5=9/640,\omega_9=1/36, \omega_{13}=23/11520;c_s^2=3/2.
\end{split}
\end{equation}
The weight matrix $\mathbf W$ is given by
\begin{equation}
\mathbf W=\text{diag}\{\omega_0,\omega_1,\omega_1,\omega_1,\omega_1,\omega_5,\omega_5,\omega_5,\omega_5,\omega_9,\omega_9,\omega_9,\omega_9,\omega_{13},\omega_{13},\omega_{13},\omega_{13}\}.
\end{equation}

(5) Case 5
\begin{equation}
\begin{split}
&\{\mathbf{c}_j,0\leq j \leq 16\}=
\left(
\begin{array}{ccccccccccccccccc}
    0 & c_1 &   0 & -c_1  &    0 & c_1 & -c_1 & -c_1 &  c_1 & 2c_1 &   0 & -2c_1  &    0 & 2c_1 & -2c_1 & -2c_1 &  2c_1  \\
    0 &   0 & c_2 &    0  & -c_2 & c_2 &  c_2 & -c_2 & -c_2 &   0 &2 c_2 &    0  & -2c_2& 2c_2 &  2c_2 & -2c_2 & -2c_2  \\
\end{array}
\right),\\
&\omega_0=0.4092905, \omega_1=0.1123018,  \omega_5=0.0335591,\omega_9=0.0017273, \omega_{13}=0.0000891;c_s^2=0.3740845.
\end{split}
\end{equation}
The weight matrix $\mathbf W$ is given by
\begin{equation}
\mathbf W=\text{diag}\{\omega_0,\omega_1,\omega_1,\omega_1,\omega_1,\omega_5,\omega_5,\omega_5,\omega_5,\omega_9,\omega_9,\omega_9,\omega_9,\omega_{13},\omega_{13},\omega_{13},\omega_{13}\}.
\end{equation}
We point out that this D2Q17 lattice model is also given by Qian \emph {et al.} \cite{Qian1998}.

In addition, there exists a special D2Q21 model that is not a special case of the D2Q25 model. In the D2Q21 model, the lattice model can be written as
\begin{equation}
\begin{split}
&\mathbf{E}_{0}=\{\mathbf{e}_j,0\leq j \leq 21\}=\\
\setlength{\arraycolsep}{1.2pt}
&\left(
\begin{array}{cccccccccccccccccccccc}
 0 &  1 &  0&  -1&  0& 1&-1&-1& 1& 2& 1&-1& -2 &-2&-1&1& 2& 3&0&-3&0\\
 0 &  0 &  1&  0& -1 & 1& 1& -1&-1&1& 2&  2& 1& -1&-2&-2&-1&0&3& 0&-3\\
 \end{array}
\right),\\
&\mathbf{E}=\operatorname{\text{diag}}(c_1, c_2) \mathbf{E}_{0}.
\end{split}
\end{equation}

Based on Eq. (\ref{eq:new11}), the weight coefficients can be determined as
\begin{subequations}
\begin{equation}
\omega_5=d_{01}d_{02}/4-8\omega_9;
\end{equation}
\begin{equation}
\omega_{17}=\left[d_{01}(3d_{01}-1)-48\omega_9\right]/144;\quad \omega_{18}=\left[d_{02}(3d_{02}-1)-48\omega_9\right]/144;
\end{equation}
\begin{equation}
\omega_1=d_{01}/2-2\omega_5-10\omega_9-9\omega_{17};\quad \omega_2=d_{02}/2-2\omega_5-10\omega_9-9\omega_{18};
\end{equation}
\begin{equation}
\omega_{0}=1-2(\omega_{1}+\omega_{2}+2\omega_5+4\omega_9+\omega_{17}+\omega_{18});
\end{equation}
\begin{equation}
W=\text{diag}\{\omega_0,\omega_1,\omega_2,\omega_1,\omega_2,\omega_5,\omega_5,\omega_5,\omega_5,\omega_9,\omega_9,\omega_9,\omega_9,\omega_9,\omega_9,\omega_9,\omega_9,\omega_{17},\omega_{18},\omega_{17},\omega_{18}\}.
\end{equation}
\end{subequations}
where $\omega_9$ is a free parameter. Let $E_{ij}$ and $M_{ij}$  be the element of the i-th row and j-th column of matrices $\mathbf E$ and $\mathbf M$, then the first $10$ rows of the transformation matrix $\mathbf M$ (i.e., the complete 0th-3rd order moments) can be expressed as
\begin{subequations}
\begin{equation}
M_{1j}=1, \quad M_{2j}=E_{1j},\quad M_{3j}=E_{2j},
\end{equation}
\begin{equation}
M_{4j}=E_{1j}E_{1j},\quad M_{5j}=E_{1j}E_{2j},\quad M_{6j}=E_{2j}E_{2j},
\end{equation}
\begin{equation}
M_{7j}=E_{1j}E_{2j}E_{2j}, \quad M_{8j}=E_{1j}E_{1j}E_{2j},  \quad M_{9j}=E_{1j}E_{1j}E_{1j},  \quad M_{10j}=E_{2j}E_{2j}E_{2j},
\end{equation}
\end{subequations}
where $j=1-21$. Rows $11-21$ in the $\mathbf M$ matrix can be designed by the researcher to make the transformation matrix invertible.

Based on Eq. (\ref{eq:CorrectQ25}), the Hermite matrix $\mathbf H$ can be written as
\begin{subequations}
\begin{equation}
H_{1j}=1, \quad H_{2j}=E_{1j},\quad H_{3j}=E_{2j},
\end{equation}
\begin{equation}
H_{4j}=E_{1j}E_{1j}-c_{s1}^2,\quad H_{5j}=E_{1j}E_{2j}, \quad H_{6j}=E_{2j}E_{2j}-c_{s2}^2,
\end{equation}
\begin{equation}
H_{7j}=E_{1j}(E_{2j}E_{2j}-c_{s2}^2), \quad H_{8j}=E_{2j}(E_{1j}E_{1j}-c_{s1}^2),
\end{equation}
\begin{equation}
H_{9j}=E_{1j}(E_{1j}E_{1j}-3c_{s1}^2)-H_{7j}a_1/b_1, \quad H_{10j}=E_{2j}(E_{2j}E_{2j}-3c_{s2}^2)-H_{8j}a_2/b_2,
\end{equation}
\end{subequations}
with
\begin{subequations}
\begin{equation}
a_1=c_1^4 c_2^2(-3d_{02}d_{01}^2+d_{02}d_{01}+48\omega_9),
\end{equation}
\begin{equation}
b_1=c_1^2 c_2^4(-d_{01}d_{02}^2+d_{01}d_{02}+48\omega_9),
\end{equation}
\begin{equation}
a_2=c_1^2 c_2^4(-3d_{01}d_{02}^2+d_{01}d_{02}+48\omega_9),
\end{equation}
\begin{equation}
b_2=c_1^4 c_2^2(-d_{02}d_{01}^2+d_{02}d_{01}+48\omega_9).
\end{equation}
\end{subequations}

\section{\label{app:sec2}Lattice models of RMRT-LB method on rectangular lattice in 3D}

In this section, we will give the rD3Q33 and rD3Q53 lattice models in the 3D case. For the 3D multispeed lattice model, only the 2-layer grid is considered in this paper for simplicity.

rD3Q53 lattice:
The lattice model used here is given by
\begin{equation}
\normalsize{
\begin{split}
&\mathbf{E}_{1}=\{\mathbf{e}_j,0\leq j \leq 53\}=\\
\setlength{\arraycolsep}{1.2pt}
&\left(
\begin{smallmatrix}
 0 &  1 & -1& 0& 0&0& 0&1&-1&-1& 1&1&-1&-1& 1&0& 0 & 0& 0&1&-1& 1&-1& 1&-1&-1& 1\\
 0 &  0 &  0& 1&-1&0& 0&1&-1& 1&-1&0& 0&  0& 0&1&-1&-1& 1&1&-1& 1&-1&-1& 1& 1&-1\\
 0 &  0 &  0& 0& 0&1&-1&0& 0& 0& 0 &1&-1& 1&-1&1&-1& 1&-1&1&-1&-1& 1& 1&-1& 1&-1\\
 \end{smallmatrix}
 \right.\\
 &\left.
 \begin{smallmatrix}
 2 & -2& 0& 0&0& 0&2&-2&-2& 2&2&-2&-2& 2&0& 0 & 0& 0&2&-2& 2&-2& 2&-2&-2& 2\\
 0 &  0& 2&-2&0& 0&2&-2& 2&-2&0& 0&  0& 0&2&-2&-2& 2&2&-2& 2&-2&-2& 2& 2&-2\\
 0 &  0& 0& 0&2&-2&0& 0& 0& 0 &2&-2& 2&-2&2&-2& 2&-2&2&-2&-2& 2& 2&-2& 2&-2\\
\end{smallmatrix}
\right),\\
&\mathbf{E}=\operatorname{\text{diag}}(c_1, c_2, c_3) \mathbf{E}_{1}.
\end{split}}
\end{equation}
To determine the weight coefficient $\omega_{j}$ for the 3D lattice model, Eq. (\ref{eq:new11f}) should be taken into account.  Based on Eq. (\ref{eq:new11}) exept Eq. (\ref{eq:new11}e), the weight coefficients can be expressed as
\begin{subequations}
\begin{equation}
\omega_1=\frac{d_{01}}{2}-2(\omega_7+\omega_{11})-4\omega_{19}-4\omega_{27}-8(\omega_{33}+\omega_{37})-16\omega_{45};
\end{equation}
\begin{equation}
\omega_2=\frac{d_{02}}{2}-2(\omega_7+\omega_{15})-4\omega_{19}-4\omega_{28}-8(\omega_{33}+\omega_{41})-16\omega_{45};
\end{equation}
\begin{equation}
\omega_3=\frac{d_{03}}{2}-2(\omega_{11}+\omega_{15})-4\omega_{19}-4\omega_{29}-8(\omega_{41}+\omega_{37})-16\omega_{45};
\end{equation}
\begin{equation}
\omega_7=\frac{d_{01}d_{02}}{4}-2\omega_{19}-16\omega_{33}-32\omega_{45};
\end{equation}
\begin{equation}
\omega_{11}=\frac{d_{01}d_{03}}{4}-2\omega_{19}-16\omega_{37}-32\omega_{45};
\end{equation}
\begin{equation}
\omega_{15}=\frac{d_{02}d_{03}}{4}-2\omega_{19}-16\omega_{41}-32\omega_{45};
\end{equation}
\begin{equation}
\omega_{19}=\frac{d_{01}d_{02}d_{03}}{8}-64\omega_{45};
\end{equation}
\begin{equation}
\omega_{27}=\frac{d_{01}(3d_{01}-1)}{24}-2(\omega_{33}+\omega_{37})-4\omega_{45};
\end{equation}
\begin{equation}
\omega_{28}=\frac{d_{02}(3d_{02}-1)}{24}-2(\omega_{33}+\omega_{41})-4\omega_{45};
\end{equation}
\begin{equation}
\omega_{29}=\frac{d_{03}(3d_{03}-1)}{24}-2(\omega_{41}+\omega_{37})-4\omega_{45};
\end{equation}
\begin{equation}
\omega_{33}=\frac{d_{01}d_{02}(3d_{01}-1)}{192}-2\omega_{45};
\end{equation}
\begin{equation}
\omega_{37}=\frac{d_{01}d_{03}(3d_{01}-1)}{192}-2\omega_{45};
\end{equation}
\begin{equation}
\omega_{41}=\frac{d_{02}d_{03}(3d_{02}-1)}{192}-2\omega_{45};
\end{equation}
\begin{equation}
\omega_{0}=1-2(\omega_{1}+\omega_{2}+\omega_{3}+\omega_{27}+\omega_{28}+\omega_{29})-4(\omega_{7}+\omega_{11}+\omega_{15}+\omega_{33}+\omega_{37}+\omega_{41})-8(\omega_{19}+\omega_{45}),
\end{equation}
\end{subequations}
where $d_{01}=\frac{c_1^2}{c_{s1}^2}$, $d_{02}=\frac{c_2^2}{c_{s2}^2}$, $d_{03}=\frac{c_3^2}{c_{s3}^2}$.
The weight matrix $\mathbf W$ is
\begin{equation}
\begin{split}
\mathbf W=&\text{diag}\{\omega_0,\omega_1,\omega_1,\omega_2,\omega_2,\omega_3,\omega_3,\omega_7,\omega_7,\omega_7,\omega_7,\omega_{11},\omega_{11},\omega_{11},\omega_{11},\omega_{15},\omega_{15},\omega_{15},\omega_{15},\\
&\omega_{19},\omega_{19},\omega_{19},\omega_{19},\omega_{19},\omega_{19},\omega_{19},\omega_{19},\omega_{27},\omega_{27},\omega_{28},\omega_{28},\omega_{29},\omega_{29},\omega_{33},\omega_{33},\omega_{33},\omega_{33},\\
&\omega_{37},\omega_{37},\omega_{37},\omega_{37},\omega_{41},\omega_{41},\omega_{41},\omega_{41},\omega_{45},\omega_{45},\omega_{45},\omega_{45},\omega_{45},\omega_{45},\omega_{45},\omega_{45}\}.
\end{split}
\end{equation}

Let $E_{ij}$ and $M_{ij}$  be the element of the i-th row and j-th column of matrices $\mathbf E$ and $\mathbf M$, then the first $20$ rows of the transformation matrix (i.e., the complete 0th-3rd order moments) can be expressed as
\begin{subequations}
\begin{equation}
M_{1j}=1, \quad M_{2j}=E_{1j},\quad M_{3j}=E_{2j},\quad M_{4j}=E_{3j},\quad M_{5j}=E_{1j}E_{1j},
\end{equation}
\begin{equation}
M_{6j}=E_{2j}E_{2j}, \quad M_{7j}=E_{3j}E_{3j}, \quad M_{8j}=E_{1j}E_{2j}, \quad M_{9j}=E_{3j}E_{1j}, \quad M_{10j}=E_{2j}E_{3j},
\end{equation}
\begin{equation}
M_{11j}=E_{1j}E_{2j}E_{2j}, \quad M_{12j}=E_{3j}E_{3j}E_{1j}, \quad M_{13j}=E_{1j}E_{1j}E_{2j},
\end{equation}
\begin{equation}
M_{14j}=E_{2j}E_{3j}E_{3j}, \quad M_{15j}=E_{3j}E_{1j}E_{1j}, \quad M_{16j}=E_{2j}E_{2j}E_{3j}, \quad M_{17j}=E_{1j}E_{2j}E_{3j},
\end{equation}
\begin{equation}
M_{18j}=E_{1j}^3, \quad M_{19j}=E_{2j}^3, \quad M_{20j}=E_{3j}^3,
\end{equation}
\end{subequations}
where $j=1-53$. Since only 0th-3rd order moments were used in the derivation process, the influence of higher-order moments (moments greater than the third order) is very small. Therefore, we only need to determine the first $20$ rows of $\mathbf M$ and Hermite  matrix $\mathbf H$, and the remaining rows can be constructed by ourselves to make the matrices invertible. The Hermite matrix $\mathbf H$ can be written as
\begin{subequations}
\begin{equation}
H_{1j}=1, \quad H_{2j}=E_{1j},\quad H_{3j}=E_{2j},\quad H_{4j}=E_{3j},\quad H_{5j}=E_{1j}E_{1j}-c_{s1}^2, \quad H_{6j}=E_{2j}E_{2j}-c_{s2}^2,
\end{equation}
\begin{equation}
H_{7j}=E_{3j}E_{3j}-c_{s3}^2, \quad H_{8j}=E_{1j}E_{2j}, \quad H_{9j}=E_{3j}E_{1j}, \quad H_{10j}=E_{2j}E_{3j},
\end{equation}
\begin{equation}
H_{11j}=E_{1j}(E_{2j}E_{2j}-c_{s2}^2), \quad H_{12j}=(E_{3j}E_{3j}-c_{s3}^2)E_{1j}, \quad H_{13j}=(E_{1j}E_{1j}-c_{s1}^2)E_{2j},
\end{equation}
\begin{equation}
H_{14j}=E_{2j}(E_{3j}E_{3j}-c_{s3}^2), \quad H_{15j}=E_{3j}(E_{1j}E_{1j}-c_{s1}^2), \quad H_{16j}=(E_{2j}E_{2j}-c_{s2}^2)E_{3j}, \quad H_{17j}=E_{1j}E_{2j}E_{3j},
\end{equation}
\begin{equation}
H_{18j}=E_{1j}^3-3c_{s1}^2 E_{1j}, \quad H_{19j}=E_{2j}^3-3c_{s2}^2 E_{2j}, \quad H_{20j}=E_{3j}^3-3c_{s3}^2 E_{3j}.
\end{equation}
\end{subequations}
As mentioned earlier, if Eq. (\ref{eq:new11e}) is not taken into account, one can also correct $\mathbf H_3=\{H_{k\cdot},k=11:20\}$ to make it block-weighted orthogonal. Here only $H_{18}=\{H_{18j}\}$, $H_{19}=\{H_{19j}\}$ and $H_{20}=\{H_{20j}\}$ need to be corrected and can be written in the following form,
\begin{subequations}
\begin{equation}
H_{18j}=E_{1j}^3-3c_{s1}^2 E_{1j}-a_1 H_{11j}-b_1 H_{12j},
\end{equation}
\begin{equation}
H_{19j}=E_{2j}^3-3c_{s2}^2 E_{2j}-a_2 H_{13j}-b_2 H_{14j},
\end{equation}
\begin{equation}
H_{20j}=E_{3j}^3-3c_{s3}^2 E_{3j}-a_3 H_{15j}-b_3 H_{16j},
\end{equation}
\end{subequations}
where $a_1$, $a_2$, $a_3$, $b_1$, $b_2$, $b_3$ are the parameters to be determined and can be given by the following equations,
\begin{subequations}
\begin{equation}
a_0=192\omega_{33}+384\omega_{45}, \quad b_0=192\omega_{37}+384\omega_{45}, \quad c_0=192\omega_{41}+384\omega_{45},
\end{equation}
\begin{equation}
a_1= \frac{\bar H_{18}\mathbf W H_{11}^{T}}{H_{11}\mathbf W H_{11}^{T}}=\frac{c_1^2}{c_2^2} \left[\frac{d_{01}d_{02}(1-3d_{01})+a_0}{d_{02}d_{01}(1-d_{02})+a_0}\right],
\end{equation}
\begin{equation}
a_2=\frac{\bar H_{19}\mathbf W H_{13}^{T}}{H_{13}\mathbf W H_{13}^{T}}= \frac{c_2^2}{c_1^2} \left[\frac{d_{02}d_{01}(1-3d_{02})+a_0}{d_{01}d_{02}(1-d_{01})+a_0}\right],
\end{equation}
\begin{equation}
a_3=\frac{\bar H_{20}\mathbf W H_{15}^{T}}{H_{15}\mathbf W H_{15}^{T}}= \frac{c_3^2}{c_1^2} \left[\frac{d_{03}d_{01}(1-3d_{03})+b_0}{d_{01}d_{03}(1-d_{01})+b_0}\right],
\end{equation}
\begin{equation}
b_1= \frac{\bar H_{18}\mathbf W H_{12}^{T}}{H_{12}\mathbf W H_{12}^{T}}=\frac{c_1^2}{c_3^2} \left[\frac{d_{01}d_{03}(1-3d_{01})+b_0}{d_{03}d_{01}(1-d_{03})+b_0}\right],
\end{equation}
\begin{equation}
b_2=\frac{\bar H_{19}\mathbf W H_{14}^{T}}{H_{14}\mathbf W H_{14}^{T}}= \frac{c_2^2}{c_3^2} \left[\frac{d_{02}d_{03}(1-3d_{02})+c_0}{d_{03}d_{02}(1-d_{03})+c_0}\right],
\end{equation}
\begin{equation}
b_3= \frac{\bar H_{20}\mathbf W H_{16}^{T}}{H_{16}\mathbf W H_{16}^{T}}=\frac{c_3^2}{c_2^2} \left[\frac{d_{03}d_{02}(1-3d_{03})+c_0}{d_{02}d_{03}(1-d_{02})+c_0}\right],
\end{equation}
\end{subequations}
where $\bar H_{18}=\{E_{1j}^3-3c_{s1}^2 E_{1j}\},~\bar H_{19}=\{E_{2j}^3-3c_{s2}^2 E_{2j}\},~H_{20}=\{E_{3j}^3-3c_{s3}^2 E_{3j}\}$.

rD3Q33 lattice:

In the rD3Q53 model, let $\omega_{33}=\omega_{37}=\omega_{41}=\omega_{45}=0$, one can obtain the rD3Q33 lattice model. Similar to the rD3Q53 lattice model, we have
\begin{equation}
\normalsize{
\begin{split}
&\mathbf{E}_{0}=\{\mathbf{e}_j,0\leq j \leq 33\}=\\
\setlength{\arraycolsep}{1.2pt}
&\left(
\begin{smallmatrix}
 0 &  1 & -1& 0& 0&0& 0&1&-1&-1& 1&1&-1&-1& 1&0& 0 & 0& 0&1&-1& 1&-1& 1&-1&-1& 1&2&-2&0& 0&0&0\\
 0 &  0 &  0& 1&-1&0& 0&1&-1& 1&-1&0& 0&  0& 0&1&-1&-1& 1&1&-1& 1&-1&-1& 1& 1&-1&0& 0&2&-2&0&0\\
 0 &  0 &  0& 0& 0&1&-1&0& 0& 0& 0 &1&-1& 1&-1&1&-1& 1&-1&1&-1&-1& 1& 1&-1& 1&-1&0& 0&0& 0&2&-2\\
 \end{smallmatrix}
\right),\\
&\mathbf{E}=\operatorname{\text{diag}}(c_1, c_2, c_3) \mathbf{E}_{0}.
\end{split}}
\end{equation}
Then the weight coefficients can be determined as
\begin{subequations}
\begin{equation}
\omega_1=\frac{d_{01}}{2}-2(\omega_7+\omega_{11})-4\omega_{19}-4\omega_{27};
\end{equation}
\begin{equation}
\omega_2=\frac{d_{02}}{2}-2(\omega_7+\omega_{15})-4\omega_{19}-4\omega_{28};
\end{equation}
\begin{equation}
\omega_3=\frac{d_{03}}{2}-2(\omega_{11}+\omega_{15})-4\omega_{19}-4\omega_{29};
\end{equation}
\begin{equation}
\omega_7=\frac{d_{01}d_{02}}{4}-2\omega_{19};
\end{equation}
\begin{equation}
\omega_{11}=\frac{d_{01}d_{03}}{4}-2\omega_{19};
\end{equation}
\begin{equation}
\omega_{15}=\frac{d_{02}d_{03}}{4}-2\omega_{19};
\end{equation}
\begin{equation}
\omega_{19}=\frac{d_{01}d_{02}d_{03}}{8};
\end{equation}
\begin{equation}
\omega_{27}=\frac{d_{01}(3d_{01}-1)}{24};
\end{equation}
\begin{equation}
\omega_{28}=\frac{d_{02}(3d_{02}-1)}{24};
\end{equation}
\begin{equation}
\omega_{29}=\frac{d_{03}(3d_{03}-1)}{24};
\end{equation}
\begin{equation}
\omega_{0}=1-2(\omega_{1}+\omega_{2}+\omega_{3}+\omega_{27}+\omega_{28}+\omega_{29})-4(\omega_{7}+\omega_{11}+\omega_{15})-8\omega_{19}.
\end{equation}
\end{subequations}
The weight matrix $\mathbf W$ is given by
\begin{equation}
\begin{split}
\mathbf W=&\text{diag}\{\omega_0,\omega_1,\omega_1,\omega_2,\omega_2,\omega_3,\omega_3,\omega_7,\omega_7,\omega_7,\omega_7,\omega_{11},\omega_{11},\omega_{11},\omega_{11},\omega_{15},\omega_{15},\omega_{15},\omega_{15},\\
&\omega_{19},\omega_{19},\omega_{19},\omega_{19},\omega_{19},\omega_{19},\omega_{19},\omega_{19},\omega_{27},\omega_{27},\omega_{28},\omega_{28},\omega_{29},\omega_{29}\}.\\
\end{split}
\end{equation}

The complete transformation matrix $\mathbf  M$ and the part of Hermite matrix $\mathbf H$ can be written as
\begin{subequations}
\begin{equation}
M_{1j}=1, \quad M_{2j}=E_{1j},\quad M_{3j}=E_{2j},\quad M_{4j}=E_{3j},\quad M_{5j}=E_{1j}E_{1j},
\end{equation}
\begin{equation}
M_{6j}=E_{2j}E_{2j}, \quad M_{7j}=E_{3j}E_{3j}, \quad M_{8j}=E_{1j}E_{2j}, \quad M_{9j}=E_{3j}E_{1j}, \quad M_{10j}=E_{2j}E_{3j},
\end{equation}
\begin{equation}
M_{11j}=E_{1j}E_{2j}E_{2j}, \quad M_{12j}=E_{3j}E_{3j}E_{1j}, \quad M_{13j}=E_{1j}E_{1j}E_{2j},
\end{equation}
\begin{equation}
M_{14j}=E_{2j}E_{3j}E_{3j}, \quad M_{15j}=E_{3j}E_{1j}E_{1j}, \quad M_{16j}=E_{2j}E_{2j}E_{3j}, \quad M_{17j}=E_{1j}E_{2j}E_{3j},
\end{equation}
\begin{equation}
M_{18j}=E_{1j}^3, \quad M_{19j}=E_{2j}^3, \quad M_{20j}=E_{3j}^3,
\end{equation}
\begin{equation}
M_{21j}=E_{1j}E_{1j}E_{2j}E_{2j}, \quad M_{22j}=E_{3j}E_{3j}E_{1j}E_{1j}, \quad M_{23j}=E_{2j}E_{2j}E_{3j}E_{3j},
\end{equation}
\begin{equation}
M_{24j}=E_{1j}E_{1j}E_{2j}E_{3j}, \quad M_{25j}=E_{1j}E_{2j}E_{2j}E_{3j}, \quad M_{26j}=E_{1j}E_{2j}E_{3j}E_{3j},
\end{equation}
\begin{equation}
M_{27j}=E_{1j}^4, \quad M_{28j}=E_{2j}^4, \quad M_{29j}=E_{3j}^4,
\end{equation}
\begin{equation}
M_{30j}=E_{1j}E_{1j}E_{2j}E_{2j}E_{3j}, \quad M_{31j}=E_{1j}E_{1j}E_{2j}E_{3j}E_{3j}, \quad M_{32j}=E_{1j}E_{2j}E_{2j}E_{3j}E_{3j},
\end{equation}
\begin{equation}
M_{33j}=E_{1j}^2E_{2j}^2E_{3j}^2.
\end{equation}
\end{subequations}

\begin{equation}\label{eq:TransM33}
\mathbf M=\mathbf D\left(
\begin{smallmatrix}
 1 &  1 &  1& 1& 1& 1& 1 & 1& 1& 1&1 &  1 &  1& 1& 1& 1& 1 & 1& 1& 1&1 &  1 &  1& 1& 1& 1& 1 & 1& 1& 1&1 & 1& 1\\
 0 &  1 & -1& 0& 0&0& 0&1&-1&-1& 1&1&-1&-1& 1&0& 0 & 0& 0&1&-1& 1&-1& 1&-1&-1& 1&2&-2&0& 0&0&0\\
 0 &  0 &  0& 1&-1&0& 0&1&-1& 1&-1&0& 0&  0& 0&1&-1&-1& 1&1&-1& 1&-1&-1& 1& 1&-1&0& 0&2&-2&0&0\\
 0 &  0 &  0& 0& 0&1&-1&0& 0& 0& 0 &1&-1& 1&-1&1&-1& 1&-1&1&-1&-1& 1& 1&-1& 1&-1&0& 0&0& 0&2&-2\\
  0 &  1 & 1& 0& 0&0& 0&1&1&1& 1&1&1&1& 1&0& 0 & 0& 0&1&1& 1&1& 1&1&1& 1&4&4&0& 0&0&0\\
  0 &  0 &  0& 1&1&0& 0&1&1& 1&1&0& 0&  0& 0&1&1&1& 1&1&1& 1&1&1& 1& 1&1&0& 0&4&4&0&0\\
 0 &  0 &  0& 0& 0&1&1&0& 0& 0& 0 &1&1& 1&1&1&1& 1&1&1&1&1& 1& 1&1& 1&1&0& 0&0& 0&4&4\\
  0 &  0 & 0& 0& 0&0& 0&1&1&-1& -1&0&0&0& 0&0& 0 & 0& 0&1&1& 1&1& -1&-1&-1& -1&0&0&0& 0&0&0\\
  0 &  0 & 0& 0& 0&0& 0&0&0&0& 0&1&1&-1& -1&0& 0 & 0& 0&1&1& -1&-1& 1&1&-1& -1&0&0&0& 0&0&0\\
  0 &  0 &  0& 0& 0&0&0&0& 0& 0& 0 &0&0& 0&0&1&1& -1&-1&1&1&-1& -1& -1&-1& 1&1&0& 0&0& 0&0&0\\
   0 &  0 & 0& 0& 0&0& 0&1&-1&-1& 1&0&0&0&0&0& 0 & 0& 0&1&-1& 1&-1& 1&-1&-1& 1&0&0&0& 0&0&0\\
   0 &  0 & 0& 0& 0&0& 0&0&0 &1& -1&-1&1&0& 0&0& 0 & 0& 0&1&-1& 1&-1& 1&-1&-1& 1&0&0&0& 0&0&0\\
   0 &  0 & 0& 0& 0&0& 0&1&-1&1& -1&0&0&0& 0&0& 0 & 0& 0&1&-1& 1&-1& -1&1&1& -1&0&0&0& 0&0&0\\
    0 &  0 & 0& 0& 0&0& 0&0&0&0& 0&0&0&0& 0&1& -1 & -1& 1&1&-1& 1&-1& -1&1&1& -1&0&0&0& 0&0&0\\
    0 &  0 & 0& 0& 0&0& 0&0&0&0& 0&1&-1&1& -1&0& 0 & 0& 0&1&-1& -1&1& 1&-1&1& -1&0&0&0& 0&0&0\\
    0 &  0 & 0& 0& 0&0& 0&0&0&0& 0&0&0&0& 0&1& -1 & 1& -1&1&-1& -1&1& 1&-1&1& -1&0&0&0& 0&0&0\\
    0 &  0 & 0& 0& 0&0& 0&0&0&0& 0&0&0&0& 0&1& -1 & 1& -1&1&-1& -1&1& 1&-1&1& -1&0&0&0& 0&0&0\\
   0 &  1 & -1& 0& 0&0& 0&1&-1&-1& 1&1&-1&-1& 1&0& 0 & 0& 0&1&-1& 1&-1& 1&-1&-1& 1&8&-8&0& 0&0&0\\
   0 &  0 &  0& 1&-1&0& 0&1&-1& 1&-1&0& 0&  0& 0&1&-1&-1& 1&1&-1& 1&-1&-1& 1& 1&-1&0& 0&8&-8&0&0\\
   0 &  0 &  0& 0& 0&1&-1&0& 0& 0& 0 &1&-1& 1&-1&1&-1& 1&-1&1&-1&-1& 1& 1&-1& 1&-1&0& 0&0& 0&8&-8\\
   0 &  0 & 0& 0& 0&0& 0&1&1&1& 1&0&0&0& 0&0& 0 & 0& 0&1&1& 1&1& 1&1&1& 1&0&0&0& 0&0&0\\
    0 &  0 &  0& 0& 0&0&0&0& 0& 0& 0 &1&1& 1&1&0&0& 0&0&1&1&1& 1& 1&1& 1&1&0& 0&0& 0&0&0\\
   0 &  0 &  0& 0& 0&0&0&0& 0& 0& 0 &0&0& 0&0&1&1& 1&1&1&1&1& 1& 1&1& 1&1&0& 0&0& 0&0&0\\
   0 &  0 &  0& 0& 0&0&0&0& 0& 0& 0 &0&0& 0&0&0&0& 0&0&1&1&-1& -1& -1&-1& 1&1&0& 0&0& 0&0&0\\
   0 &  0 &  0& 0& 0&0&0&0& 0& 0& 0 &0&0& 0&0&0&0& 0&0&1&1&-1& -1& 1&1& -1&-1&0& 0&0& 0&0&0\\
   0 &  0 &  0& 0& 0&0&0&0& 0& 0& 0 &0&0& 0&0&0&0& 0&0&1&1&1& 1& -1&-1& -1&-1&0& 0&0& 0&0&0\\
   0 &  1 & 1& 0& 0&0& 0&1&1&1& 1&1&1&1& 1&0& 0 & 0& 0&1&1& 1&1& 1&1&1& 1&16&16&0& 0&0&0\\
   0 &  0 &  0& 1&1&0& 0&1&1& 1&1&0& 0&  0& 0&1&1&1& 1&1&1& 1&1&1& 1& 1&1&0& 0&16&16&0&0\\
   0 &  0 &  0& 0& 0&1&1&0& 0& 0& 0 &1&1& 1&1&1&1& 1&1&1&1&1& 1& 1&1& 1&1&0& 0&0& 0&16&16\\
   0 &  0 &  0& 0& 0&0&0&0& 0& 0& 0 &0&0& 0&0&0&0& 0&0&1&-1&-1& 1& 1&-1& 1&-1&0& 0&0& 0&0&0\\
   0 &  0 &  0& 0& 0&0&0&0& 0& 0& 0 &0&0& 0&0&0&0& 0&0&1&-1&1& -1& -1&1& 1&-1&0& 0&0& 0&0&0\\
   0 &  0 &  0& 0& 0&0&0&0& 0& 0& 0 &0&0& 0&0&0&0& 0&0&1&-1&1& -1& 1&-1& -1&1&0& 0&0& 0&0&0\\
   0 &  0 &  0& 0& 0&0&0&0& 0& 0& 0 &0&0& 0&0&0&0& 0&0&1&1&1& 1& 1&1& 1&1&0& 0&0& 0&0&0\\
 \end{smallmatrix}
\right),\\
\end{equation}
where
\begin{subequations}\label{eq:TransM33-D}
\begin{equation}
\mathbf D=\text{diag}\{\mathbf D_1,\mathbf D_2,\mathbf D_3\},
\end{equation}
\begin{equation}
\mathbf D_1=\text{diag}\{1,c_1,c_2,c_3,c_1^2,c_2^2,c_3^2,c_1c_2,c_3c_1,c_2c_3\},
\end{equation}
\begin{equation}
\mathbf D_2=\text{diag}\{c_1c_2^2,c_1c_3^2,c_2c_1^2,c_2c_3^2,c_3c_1^2,c_3c_2^2,c_1c_2c_3,c_1^3,c_2^3,c_3^3\},
\end{equation}
\begin{equation}
\mathbf D_3=\text{diag}\{c_1^2c_2^2,c_1^2c_3^2,c_2^2c_3^2,c_1^2c_2c_3,c_1c_2^2c_3,c_1c_2c_3^2,c_1^4,c_2^4,c_3^4,c_1^2c_2^2c_3,c_1^2c_2c_3^2,c_1c_2^2c_3^2,c_1^2c_2^2c_3^2\}.
\end{equation}
\end{subequations}

\begin{subequations}
\begin{equation}
H_{1j}=1, \quad H_{2j}=E_{1j}, \quad H_{3j}=E_{2j},\quad H_{4j}=E_{3j}, \quad H_{5j}=E_{1j}E_{1j}-c_{s1}^2, \quad H_{6j}=E_{2j}E_{2j}-c_{s2}^2,
\end{equation}
\begin{equation}
H_{7j}=E_{3j}E_{3j}-c_{s3}^2, \quad H_{8j}=E_{1j}E_{2j}, \quad H_{9j}=E_{3j}E_{1j}, \quad H_{10j}=E_{2j}E_{3j},
\end{equation}
\begin{equation}
H_{11j}=E_{1j}(E_{2j}E_{2j}-c_{s2}^2), \quad H_{12j}=(E_{3j}E_{3j}-c_{s3}^2)E_{1j}, \quad H_{13j}=(E_{1j}E_{1j}-c_{s1}^2)E_{2j},
\end{equation}
\begin{equation}
H_{14j}=E_{2j}(E_{3j}E_{3j}-c_{s3}^2), \quad H_{15j}=E_{3j}(E_{1j}E_{1j}-c_{s1}^2), \quad H_{16j}=(E_{2j}E_{2j}-c_{s2}^2)E_{3j}, \quad H_{17j}=E_{1j}E_{2j}E_{3j},
\end{equation}
\begin{equation}
H_{18j}=E_{1j}^3-3c_{s1}^2 E_{1j}-a_1\left( H_{11j}/b_2+ H_{12j}/b_3\right),
\end{equation}
\begin{equation}
H_{19j}=E_{2j}^3-3c_{s2}^2 E_{2j}-a_2\left( H_{13j}/b_1+ H_{14j}/b_3\right),
\end{equation}
\begin{equation}
H_{20j}=E_{3j}^3-3c_{s3}^2 E_{3j}-a_3\left( H_{15j}/b_1+ H_{16j}/b_2\right),
\end{equation}
\end{subequations}
where $a_1=3c_{s1}^2-c_1^2$, $a_2=3c_{s2}^2-c_2^2$, $a_3=3c_{s3}^2-c_3^2$, $b_1=c_{s1}^2-c_1^2$, $b_2=c_{s2}^2-c_2^2$, $b_3=c_{s3}^2-c_3^2$.

\end{document}